\documentclass[twoside,twocolumn,9pt]{article}
\usepackage{extsizes}
\usepackage[super,sort&compress,comma]{natbib} 
\usepackage[version=3]{mhchem}
\usepackage[left=1.5cm, right=1.5cm, top=1.785cm, bottom=2.0cm]{geometry}
\usepackage{balance}
\usepackage{bm}
\usepackage{mathptmx}
\usepackage{sectsty}
\usepackage{graphicx}
\usepackage{caption}
\usepackage{subcaption}
\usepackage{lastpage}
\usepackage[format=plain,justification=justified,singlelinecheck=false,font={stretch=1.125,small,sf},labelfont=bf,labelsep=space]{caption}
\usepackage{float}
\usepackage{fancyhdr}
\usepackage{fnpos}
\usepackage[english]{babel}
\addto{\captionsenglish}{%
  
}
\usepackage{array}
\usepackage{droidsans}
\usepackage{charter}
\usepackage[T1]{fontenc}
\usepackage[usenames,dvipsnames]{xcolor}
\usepackage{setspace}
\usepackage[compact]{titlesec}
\usepackage{hyperref}

\usepackage{epstopdf}

\definecolor{cream}{RGB}{222,217,201}

\begin{document}

\pagestyle{fancy}
\thispagestyle{plain}
\fancypagestyle{plain}{
\renewcommand{\headrulewidth}{0pt}
}

\makeFNbottom
\makeatletter
\renewcommand\LARGE{\@setfontsize\LARGE{15pt}{17}}
\renewcommand\Large{\@setfontsize\Large{12pt}{14}}
\renewcommand\large{\@setfontsize\large{10pt}{12}}
\renewcommand\footnotesize{\@setfontsize\footnotesize{7pt}{10}}
\makeatother

\renewcommand{\thefootnote}{\fnsymbol{footnote}}
\renewcommand\footnoterule{\vspace*{1pt}%
\color{cream}\hrule width 3.5in height 0.4pt \color{black}\vspace*{5pt}} 
\setcounter{secnumdepth}{5}

\makeatletter 
\renewcommand\@biblabel[1]{#1}            
\renewcommand\@makefntext[1]%
{\noindent\makebox[0pt][r]{\@thefnmark\,}#1}
\makeatother 
\renewcommand{\figurename}{\small{Fig.}~}
\sectionfont{\sffamily\Large}
\subsectionfont{\normalsize}
\subsubsectionfont{\bf}
\setstretch{1.125} 
\setlength{\skip\footins}{0.8cm}
\setlength{\footnotesep}{0.25cm}
\setlength{\jot}{10pt}
\titlespacing*{\section}{0pt}{4pt}{4pt}
\titlespacing*{\subsection}{0pt}{15pt}{1pt}

\fancyfoot{}
\fancyfoot[LO,RE]{\vspace{-7.1pt}}
\fancyfoot[CO]{\vspace{-7.1pt}\hspace{13.2cm}}
\fancyfoot[CE]{\vspace{-7.2pt}\hspace{-14.2cm}}
\fancyfoot[RO]{\footnotesize{\sffamily{ ~\textbar  \hspace{2pt}\thepage}}}
\fancyfoot[LE]{\footnotesize{\sffamily{~\textbar  \hspace{2pt}\thepage}}}
\fancyhead{}
\renewcommand{\headrulewidth}{0pt} 
\renewcommand{\footrulewidth}{0pt}
\setlength{\arrayrulewidth}{1pt}
\setlength{\columnsep}{6.5mm}
\setlength\bibsep{1pt}

\makeatletter 
\newlength{\figrulesep} 
\setlength{\figrulesep}{0.5\textfloatsep} 

\newcommand{\topfigrule}{\vspace*{-1pt}%
\noindent{\color{cream}\rule[-\figrulesep]{\columnwidth}{1.5pt}} }

\newcommand{\botfigrule}{\vspace*{-2pt}%
\noindent{\color{cream}\rule[\figrulesep]{\columnwidth}{1.5pt}} }

\newcommand{\dblfigrule}{\vspace*{-1pt}%
\noindent{\color{cream}\rule[-\figrulesep]{\textwidth}{1.5pt}} }

\makeatother

\twocolumn[
\sffamily

\begin{center}
    \LARGE{\textbf{Active nematics in corrugated channels}}
    \\\vspace{0.2cm}
    \large{Jaideep P. Vaidya,$^{\ast}$\textit{$^{a}$} Tyler N. Shendruk,\textit{$^{b}$} and Sumesh P. Thampi\textit{$^{a}$}}
\end{center}

 \noindent\normalsize{Active nematic fluids exhibit complex dynamics in both bulk and in simple confining geometries. However, complex confining geometries could have substantial impact on active spontaneous flows. Using multiparticle collision dynamics simulations adapted for active nematic particles, we study the dynamic behaviour of an active nematic fluid confined in a corrugated channel. The transition from a quiescent state to a spontaneous flow state occurs from a weak swirling flow  to a strong coherent flow due to the presence of curved-wall induced active flows. We show that active nematic fluid flows in corrugated channels can be understood in two different ways: (i) as the result of an early or delayed flow transition when compared with that in a flat-walled channel of appropriate width and (ii) boundary-induced active flows in the corrugations providing an effective slip velocity to the coherent flows in the bulk. Thus, our work illustrates the crucial role of corrugations of the confining boundary in dictating the flow transition and flow states of active fluids. } \\


\vspace{0.6cm}
  ]

\renewcommand*\rmdefault{bch}\normalfont\upshape
\rmfamily
\section*{}
\vspace{-1cm}


\footnotetext{\textit{$^{a}$~Department of Chemical Engineering, Indian Institute of Technology Madras, Chennai, 600036, India; E-mail: sumesh@iitm.ac.in}}
\footnotetext{\textit{$^{b}$~School of Physics and Astronomy, The University of Edinburgh, Peter Guthrie Tait Road, Edinburgh, EH9 3FD, United Kingdom}}




\section{Introduction}
Active and living systems are out of equilibrium due to continuous influx of energy at microscopic scales. Common examples of these systems are suspensions of cytoskeletal filaments powered by molecular motors \cite{martinez2019selection, sanchez2012spontaneous, chandrakar2020confinement,bhaskaranInstability}, bacterial suspensions \cite{wensink2012meso, liu2021viscoelastic}, cellular layers \cite{saw2017topological,armengol2023epithelia,hallatschek2023proliferating} and Janus catalysts \cite{lin2018collective, popescu2020chemically,wang2022forces}. Continuous energy injection not only leads to motion of individual active particles but also to rather exciting phenomenon of collective motion. Collective dynamics in dense suspensions of active particles usually consist of jets and circulations \textemdash~a state commonly called active turbulence\cite{doostmohammadi2018active,giomiturbulence,activeturbulence}. Several experimental realizations of active turbulence have been demonstrated in literature \cite{sanchez2012spontaneous, wensink2012meso,gupta2022active,arora2022motile,alert2022active, aranson2022bacterial}. 

Active nematics are model fluids that capture the dynamics of some of the above mentioned systems \cite{balasubramaniam2022active, thampi2022channel, doostmohammadi2018active,giomi2015geometry}. By incorporating the active stress generated by microscopic active entities, active nematics build upon the theory of passive nematic liquid crystals \cite{simha2002hydrodynamic, marchetti2013hydrodynamics, koch2011collective}. Hence, mechanisms, such as generation and dynamics of topological defects that sustain active turbulence, are inherent features of active nematic fluids \cite{thampi2014instabilities, thampi2016active, giomi2014defect, doostmohammadi2018active, Head2024}. Active stress generates hydrodynamic instabilities overcoming the elastic, viscous and frictional forces in the active fluid. In a confined active nematic fluid, the competition between activity and the opposing forces gives rise to a spontaneous flow transition at a critical activity \cite{R.Voituriez_2005, thampi2022channel, Abhik, Duclos2018, chandrakar2020confinement}. 

Confinements play a major role in dictating the dynamic state of active systems, much different from those of driven systems. For example, when active nematics are confined in microchannels of square cross section, they exhibit a multitude of states, such as a state of no fluid flow, streaming states like unidirectional, oscillatory or double helix flows and swirling states including vortex lattices or turbulent flows. All these depend on varying a single parameter, namely the ratio of the channel width to the active length scale \cite{thampi2022channel, Dogic, D.Marenduzzo}. Despite the crucial importance boundaries play a role in dictating the dynamics of active fluids, most investigations so far have considered only channels with flat walls \cite{shendruk2017dancing, Santhan3D, chandrakar2020confinement, Duclos2018, Abhik, wioland2016directed,bhaskaran3D}, cylindrical channels \cite{ravnik2013confined}, or annular rings \cite{Dogic, joshi2023disks,chen2018dynamics,Bhaskarandisks}. The effect of non-uniformity of the channel walls on the dynamics of active fluids has not been addressed in the literature yet.

However, some of the previous investigations have hinted at the effect of non-uniform boundary walls on the dynamics of active nematic fluids. In the experiments of \citet{Dogic}, asymmetric corrugations of the channel walls were used to direct streaming flows generated by mixtures of microtubule bundles and kinesin molecular motors. When the active suspension was confined in a toroid of square cross section it produced a streaming flow in the channel, with equal probability of streaming in clockwise and counter clockwise directions. On the other hand, when the channel walls of the toroid were modified by providing saw-tooth like notches, it was possible to control the directionality of the streaming flows. Such a control on the directionality of active fluid flows is essential for any future engineering applications. While the exact mechanism that leads to the directionality is not clear, it was shown that the directionality of the flow in the toroid was accompanied by swirls in the undulated sections of the channel wall. Being at low Reynolds number, the presence of swirling flows in the undulations of the channel is also intriguing. Asymmetric structures driving directionality has also been demonstrated in other active systems \cite{bechinger2016active}. Another feature the curved boundaries provide is their capability to drive active fluid flows. The boundary curvature driven flows arise since active entities preferentially orient at the solid wall, as demonstrated using theoretical \cite{Joanny, houston2023colloids}, computational \cite{thampi2016advance} and experimental \cite{ray2023rectified} tools. However, the role of these boundary-induced active flows in dictating the flow transition, the directionality of flow or the flow state itself has not yet been analyzed. 

In this work, we computationally investigate the flow generated by active nematic fluids when confined in corrugated channels. Since the spontaneous flow transition in an active nematic fluid is well studied theoretically \cite{R.Voituriez_2005}, experimentally \cite{Duclos2018, chandrakar2020confinement} and numerically \cite{doostmohammadi2018active, thampi2022channel}, we primarily focus on the effect of corrugations on the flow transition from a no-flow state to a flow state of an active nematic fluid. The resulting flow state immediately after the transition is, typically, a streaming flow state and analysis of this flow further allows us to isolate the role of corrugations and boundary-induced active flows. The investigation is performed using a numerical framework based on multiparticle collision dynamics \cite{Tyler2022}. 

In section \ref{sec:method}, we discuss the system and the multiparticle collision dynamics algorithm for active nematic fluids. We first demonstrate the validity of the algorithm illustrating the spontaneous flow transition in flat-walled channels (section~\ref{sec:results}). This is followed by an analysis of flow states generated in corrugated channels and contrasted with those in a flat-walled channel. Further, the role of geometry in establishing the flow transition and streaming-flow state is characterised and generalised. Finally, we provide a slip-based mechanism to understand the spontaneous flow transition in a corrugated channel and conclude the work (section~\ref{sec:conclusion}).



\section{Methodology}
\label{sec:method}
\subsection{Active nematic multiparticle collision dynamics (AN-MPCD)}
The active nematic fluid confined in a corrugated channel is simulated using an active-nematic multiparticle collision dynamics (AN-MPCD) algorithm. The AN-MPCD is a particle based, mesoscopic algorithm proven to be useful to simulate various soft matter systems. By assigning a nematic stress to the particles, AN-MPCD algorithm simulates the dynamics of active nematic fluids\cite{Tyler2022,humberto,Ryanprl,kozhukhov2024mitigating}. AN-MPCD algorithm discretizes the continuum into \(N\) point-like particles, each with mass \(m\), position \(\mathbf{r}_i\), velocity \(\mathbf{v}_i\) and orientation \(\mathbf{u}_i\). The algorithm comprises of two steps: streaming and collision. During streaming, each particle moves ballistically for a time \(\delta{t}\) to new position
\begin{equation}
    \mathbf{r}_i(t+\delta{t})=\mathbf{r}_i(t) + \mathbf{v}_i(t)\delta{t}.
\end{equation} 
The collisions are accomplished by partitioning the domain into cubic cells. These collisions are governed by mesoscopic collision operator \(\boldsymbol{\Xi_{i}}\), which is stochastic and ensures the conservation of  momentum within each cell. This momentum transfer among particles, thereby updates the velocities according to
\begin{equation}
    \mathbf{v}_i(t+\delta{t}) = \mathbf{v}(t) + \boldsymbol{\Xi_{i}}
\end{equation}
where  \(\mathbf{v}(t) = \langle \mathbf{v}_i \rangle\) is the center of mass velocity of cell, calculated at position \(\mathbf{r}\) in the cell. Since all particles in the cell are assumed to have identical mass \(m\), the center of mass velocity of the particles in the cell remains unchanged during the collisions process. Further, the conservation of energy and angular momentum are also achieved by the constrained stochastic exchange of particle velocities. The hydrodynamic velocity field can be obtained as the average velocity \(\mathbf{v}(t)\) at position \(\mathbf{r}\) in the cell. 

The collision operator injects energy into the system, corresponding to an extensile or contractile force dipole density. This is achieved by incorporating the local active stress proportional to the nematic tensor order parameter \(\mathbf{Q}\), such that the force dipole coaligns with the local director field \(\mathbf{n}\) of the cell. Thus, the collision operator is a linear combination of passive and active contributions,
\begin{equation}
    \boldsymbol{\Xi}_{i} = \boldsymbol{\Xi}_{i}^N + a\delta{t}\left(\frac{\kappa_i - \left\langle\kappa_j\right\rangle }{m}\right) \mathbf{n}
\end{equation}
where, \(\boldsymbol{\Xi}_{i}^N\) is a nematic multi-particle collision operator (see Appendix \ref{Appendix A}) which corresponds to the passive contribution\cite{Tyler2022,Tyler2015} arising from the orientational order of nematic particles. The active contribution comprises of two terms: (i) individual impulses (per unit mass) \((a\delta{t}/m)\kappa_i\mathbf{n}\) for each particle \(i\), representing the active force driving a change in momentum at each time step, and  (ii) a term to ensure local conservation of momentum \(-(a\delta{t}/m)\langle\kappa_j\rangle\mathbf{n}\), which is averaged over a cell. The first term  \((a\delta{t}/m)\kappa_i\mathbf{n}\) is composed of three factors: (i) \(a\), which represents the local active dipole strength in cell given by \(a = \Tilde{N}\times\alpha\) where $\alpha$ is the activity of each particle and $\Tilde{N}$ is the number of particles in a cell, (ii) \(\kappa_i\mathbf{n} = \pm\mathbf{n}\) sets the direction of the active force acting on the particle \(i\), and (iii) \(\delta{t}/m\) ensuring that \(a\) has units of force. As stated above, \(a\) is directly proportional to the number of local active agents in the cell, and positive (negative) particle activity \(\alpha\) leads to extensile (contractile) active nematics. The factor \(\kappa_i\) is a parallel/antiparallel coefficient. For particles that are above the plane defined by the center of mass \(\mathbf{r}\) and the director director \(\mathbf{n}\), \(\kappa_i = +1\) indicating that the particle is driven "forward" and particles below the plane are kicked  "backward" \((\kappa_i = -1)\).

The collision operator also updates particles orientation \(\mathbf{u}_i\), thus computing the dynamics of nematic tensor order parameter, defined as, \(\mathbf{Q} = \left\langle 2\mathbf{u}_i\mathbf{u}_i-\mathbf{\hat{1}}\right\rangle\). Here \(\mathbf{\hat{1}}\) denotes the identity matrix. The tensor order parameter \(\mathbf{Q}\) quantifies the extent and direction of orientational order through its largest eigenvalue $S$ and the corresponding eigenvector \(\mathbf{n}\) in cell. The orientations are updated based on the local equilibrium distribution of the orientation field 
\begin{equation}
    \mathbf{u}_i(t+\delta{t}) = \mathbf{n}(t)+ \boldsymbol{\eta}_i
    \label{eqn:orientation collision}
\end{equation}
where, the noise \(\boldsymbol{\eta}_i\) is drawn from the Maier-Saupe distribution \(\sim\exp( (U S/k_{B}T)[\mathbf{u}_i \cdot\mathbf{n}]^2)\). The mean-field interaction constant \(U\) and inverse thermal energy \(1/k_{B}T\) govern the width of the distribution around \(\mathbf{n}(t)\). Consequently, when \(U S/k_{B}T\) is small, all orientations are equally probable, leading to an isotropic phase. For large \(U S/k_{B}T\), the distribution sharply centers about \(\mathbf{n}\) resulting in a nematic phase. The gradients in velocity are coupled to shear alignment through discretised Jeffery's equation (see Appendix \ref{Appendix A}) of slender rod with tumbling parameter $\lambda$\cite{Tyler2015}. The value of the tumbling parameter sets the nematic fluid in either the shear-aligning or flow-tumbling regimes. Previous investigations\cite{Tyler2015,Tyler2022} have shown that when $\lambda <1$ the nematogens tumble with the flow. A relaxation parameter $\chi$, allowing averaging of Jeffery's equation over a small number of time steps and a viscous rotation coefficient, $\gamma_R$ are incorporated in the \(\boldsymbol{\Xi}_{i}^N\) (see Appendix \ref{Appendix A}) for the coupling of velocity field to director dynamics. The coupling in necessary to address the backflow effects produced due to the torques on the MPCD particles. Thus, the fluid velocity and orientation of the nematic fluid is two-way coupled to produce active nematohydrodynamics.

\begin{figure}[]
 \centering
 \includegraphics[scale = 0.6]{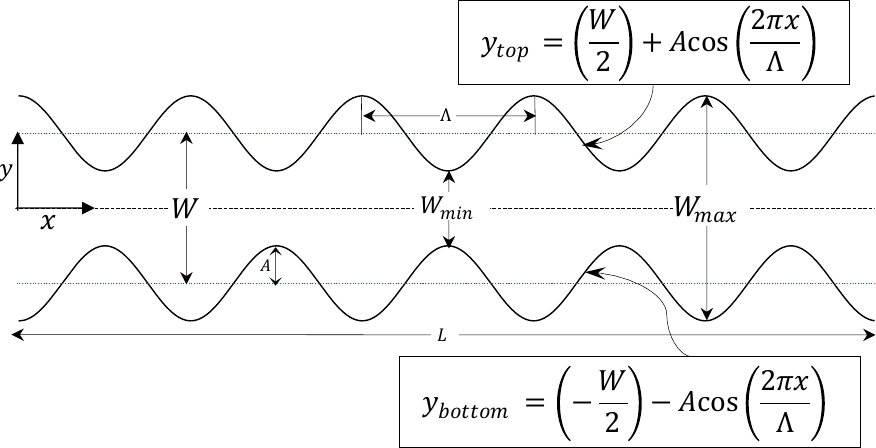}
\caption{Schematic diagram of the corrugated channel used in the simulations: \(A\) and \(\Lambda\) are amplitude and wavelength of the corrugations, respectively, \(L\) is the length, \(W\) is the mean width, \(W_{min}\) is the minimum width and $W_{max}$ is the maximum width of the channel. The boxed equations in the figure describe the geometry of the top and bottom wall of the channel.}
 \label{fgr:Schematic}
\end{figure}

\subsection{Simulation details}
\label{2.2}
We consider flow aligning nematics with extensile activity \((\alpha > 0)\) confined in a corrugated channel\cite{Wamsler2024}. A schematic of the corrugated channel used for the simulations is shown in Fig.~\ref{fgr:Schematic}. Cartesian coordinates are adopted \(-\) \(x\) along the length of the channel, and \(y\) in the perpendicular direction, i.e., across the channel and channel centerline is at $y=0$. The profile of the corrugated walls in the channel is represented by cosine functions and the equations describing the top and bottom wall are: \(y_{\text{top}}(x) = (W/2) + A \cos \left(2\pi x/\Lambda\right) \), and \(y_{\text{bottom}}(x) = (-W/2) - A \cos \left(2\pi x/\Lambda\right) \), respectively. Here, \(W\) is the mean width of the channel and \(A\) and \(\Lambda\) are the amplitude and  wavelength of the corrugations respectively. Since the width varies along the channel, the minimum width, \(W_{\text{min}}\), is defined as the distance between the trough of the top wall, \(W/2 - A\), and the crest of the bottom wall, \(-W/2 + A\), \textit{i.e.,} \(W_{\text{min}} = W - 2A\). Similarly the maximum width in the corrugated channel is, $W_{max} = W + 2A$.

As discussed in section 2.1, the total activity of the fluid in the channel is proportional to the number of active particles in the system. Therefore, for a fair comparison, (i) it is necessary for the volume of the fluid in the corrugated and flat channel to be identical and (ii) the fluid is of same density and viscosity in both the cases. In this work, two-dimensional simulations are considered; therefore, the area, of the corrugated channel occupied by the active fluid (volume per unit length), is given by:
\begin{equation}
    \text{Area}= \int_{0}^{L} \left(y_{top}(x) - y_{bottom}(x)\right)dx = W\times{L}
\end{equation}
which is equal to area of the flat-walled channel. This equivalence ensures the number of active particles used to discretize the continuum in both flat and corrugated channels are equal. Secondly, the density and viscosity of the fluid in the AN-MPCD algorithm is determined by the average particle density of the cell\cite{Gompper2009}. Therefore, the average particle density of the cell is set to \(\left\langle{N}\right\rangle = 20\) in both cases, to ensure same fluid properties in all channel geometries.

Results are reported in MPCD units of cell size $l = 1$, particle mass $m = 1$ and thermal energy $k_\text{B}T = 1$, which leads to units of time $\tau = l \sqrt{m/k_\text{B} T} =1$. The AN-MPCD particles are initialised with random velocities and oriented along the \(y\) axis. In the simulations, the mean-field interaction constant \(U=10\) is set to achieve the nematic phase\cite{Tyler2022}. The tumbling parameter \(\lambda = 2\) is chosen for the active nematic fluid to be in shear-aligning regime\cite{Tyler2022}. The heuristic coupling coefficient, $\chi$ is fixed to 0.5  and the viscous rotation coefficient is fixed as $\gamma_R=0.01$. The time step is set to \(\delta{t} = 0.1\) and simulations are run for a total of \(6\times10^5\) time steps which includes a warm-up phase of \(3\times10^5\) time steps.  The channels investigated are of the following dimensions: (i) a flat-walled channel (\(A=0\)) with length \(L=100\), width \(W=20\) and (ii) a corrugated channel with wavelength,  \(\Lambda=20\), amplitude \(A=3\) and same length and width as in (i) unless specified otherwise. Periodic boundary conditions are imposed along the \(x\) direction for the flow velocity and orientation tensor (see Appendix \ref{sctn:PBC}). The walls of the channel are impermeable with no-slip boundary conditions using phantom particles (see Appendix \ref{sctn:walls}). A strong homeotropic boundary condition is enforced for the director field ensuring that the director field is oriented perpendicular to the channel walls (see Appendix \ref{sctn:rules}).

\subsection{Corrugated channel boundaries}
The boundary conditions that simulate the active nematic fluid in a corrugated channel as shown in Fig.~\ref{fgr:Schematic}, comprises of two parts: (i) the wall surface and (ii) the boundary rules for the transformation of the velocity during the collision event. The channel walls are sinusoid functions as described in Appendix~\ref{Appendix B}. The rules for velocity transformation during the collision event with a wall are bounce-back. The transformation of the orientation of the particle involves multiple operations to obtain desired anchoring on the curved wall. Details of implementation of these boundary conditions are provided in Appendix~\ref{sctn:BC}, and is closely adapted from~\citet{Wamsler2024}.

\begin{figure*}[t!h!]
     \centering
    \includegraphics[width = 15cm]{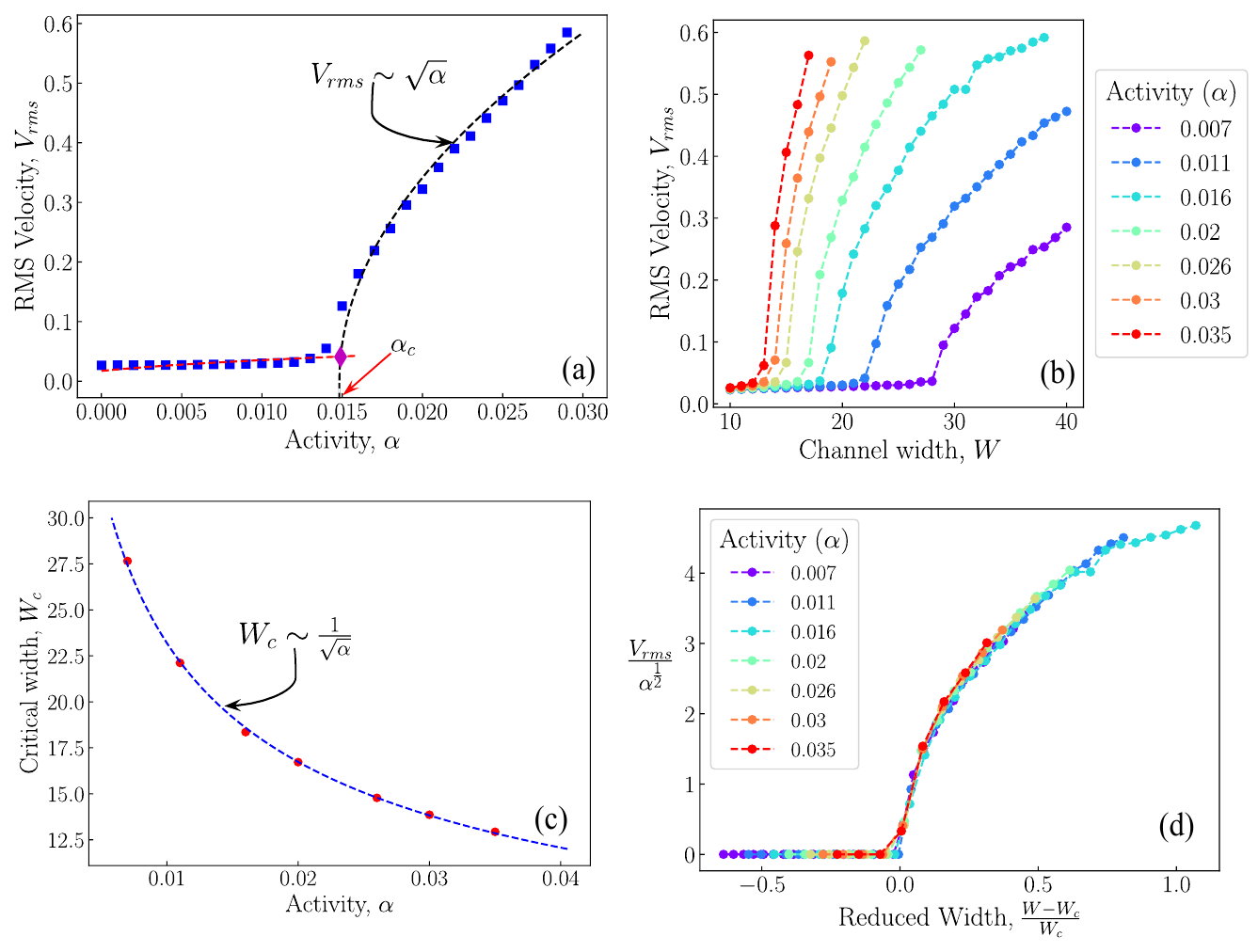}
    \caption{Spontaneous flow transition in a flat-walled channel characterized by change in root-mean-squared (RMS) velocity \(V_{rms}\) as a function of (a) activity \(\alpha\) for \(W = 20\) and (b) channel width \(W\) for various activities. (c) The critical width \(W_c\) plotted as a function of activity for a flat-walled channel. (d) Plot of \(V_{rms}/\alpha^{1/2}\) as a function of the reduced distance from the critical point \((W-W_c)/W_c\) for various activities, collapsing the data reported in figure (b) into a single curve.}
    \label{fgr:fig2}
\end{figure*}

\section{Results and Discussion}
\label{sec:results}
 We now present the results of two dimensional simulations of an active nematic fluid confined in flat and corrugated channels. We first investigate the active nematic fluid flow in a flat-walled channel by systematically varying the activity \(\alpha\) and channel width \(W\). The results confirm that AN-MPCD simulations are in agreement with previous studies \cite{D.Marenduzzo,Giomi_2012,R.Voituriez_2005} (section~\ref{3.1}-\ref{3.2}). Subsequently, by replacing the flat walls with corrugated walls, we identify the differences in the fluid flow behaviour and underlying physical mechanisms involved (section~\ref{3.3}-\ref{3.4}). The amplitude, \(A\) and the wavelength, \(\Lambda\) of the corrugations are altered to understand the effect of geometrical parameters. Lastly, the influence of circulations trapped in the corrugations are discussed along with the results of a linear stability analysis where the effect of the swirls trapped in the corrugations of the channel is interpreted as an effective slip velocity (section~\ref{3.5}). 

\subsection{Spontaneous flow transition in a flat-walled channel}
\label{3.1}
Theoretical\cite{D.Marenduzzo,Giomi_2012,R.Voituriez_2005} and experimental\cite{Duclos2018} investigations have shown that confined active nematic fluid undergoes a spontaneous flow transition at a threshold critical activity\cite{R.Voituriez_2005}, herein denoted as \(\alpha_c\). This flow transition can be characterised by a non-flowing state when \(\alpha<\alpha_c\), and a spontaneous flow state when \(\alpha>\alpha_c\). To test this,  using the AN-MPCD algorithm, we simulate an active nematic fluid confined in a flat-walled channel with systematic increments in the value of activity. Both unidirectional and bidirectional flows have been discussed in the literature following a spontaneous flow transition \cite{doostmohammadi2018active,Duclos2018,Giomi_2012,Li2021,D.Marenduzzo,Pratley2024,R.Voituriez_2005}; however, in this manuscript we focus on unidirectional flows and the scaling\cite{,R.Voituriez_2005,Giomi_2012} relationships that connect the strength of the flow, activity and confinement length scale. The flow in the channel is characterized by the root-mean-squared velocity, \(V_{rms} = \left\langle{\sqrt{\langle{v^2}\rangle}}\right\rangle_t\), where \(v\) is the fluid velocity in cell. The calculation of \(V_{rms}\) is a multi-step process; first the squared fluid velocities $v^2$ is computed within each cell and spatially averaged to obtain \(\langle{v^2}\rangle(t)\). Subsequently, a steady-state is identified and the time average of \(\sqrt{\langle{v^2}\rangle}\) is computed. It is evident that a no-flow state (negligible \(V_{rms}\)) is observed for \(\alpha<\alpha_c\) and a flow state is observed for \(\alpha>\alpha_c\) (Fig.~\ref{fgr:fig2}(a)). Above $\alpha_c$, \(V_{rms}\) continues to increase as a function of activity. The sudden change in \(V_{rms}\) is indicative of phase transition-like behavior. We determine the critical activity by curve fitting the \(V_{rms}\) data of both no-flow and flow states to the mathematical form \(V_{rms} \propto \sqrt{\alpha}\) (dashed curves in Fig.~\ref{fgr:fig2}(a)). The critical activity \(\alpha_c\) of the system is identified from the intersection point (diamond in Fig.~\ref{fgr:fig2}(a)) of the fitted curves.

The critical activity for spontaneous flow transition of confined active nematics can be calculated from a linear stability analysis of the governing continuum equations \cite{Giomi_2012} to be of the form
\begin{equation}
    \Hat{\alpha}_c = \frac{8 \pi^2 \mu \gamma^{-1} K}{W^2 S (2+\Hat{\lambda})}
    \label{eqn crtitcal activity flat}
\end{equation}
where, \(\gamma\) is the rotational viscosity, \(K\) is the Frank's elasticity constant, \(\mu\) is the fluid viscosity and $\Hat{\lambda}$ is the flow aligning parameter. We use $\Hat{\alpha}_c$ and $\Hat{\lambda}$ for critical activity and flow aligning parameter in continuum theory to differentiate the parameters from their AN-MPCD counterparts. From Eq.~\ref{eqn crtitcal activity flat}, it is apparent that apart from the fluid properties,  critical activity is also a function of the channel width, $W$. This indicates the presence of a critical width \(W_c\) that is a function activity.

To quantify spontaneous flow transition using \(W_c\), we systematically increase the width of the channel for various activities (Fig.~\ref{fgr:fig2}(b)), where \(V_{rms}\) is plotted as a function of channel width \(W\). The active nematic fluid undergoes the transition from no-flow to spontaneous flow as the channel width is increased beyond $W_c$ for a specified activity. The critical width is computed using a similar methodology to determining critical activity by curve fitting the data using  \(V_{rms} \propto \sqrt{W - W_c}\), and extracting \(W_c\). The critical width $W_c$ decreases as activity increases (Fig.~\ref{fgr:fig2}(c)). This analysis confirms that for a flat-walled channel, the critical width scales as \(W_c \sim \frac{1}{\sqrt{\alpha}}\), consistent with Eq.~\ref{eqn crtitcal activity flat} and previous investigations \cite{R.Voituriez_2005}.

Further, it can be seen that the dependence of \(V_{rms}\) on \(W\) is similar for different activities (Fig.~\ref{fgr:fig2}(b)). Therefore, we collapse the data by plotting \(V_{rms}/\alpha^{1/2}\) as a function of \((W-W_c)/W_c\) (Fig~\ref{fgr:fig2}(d)). This collapse is explained through a scaling analysis from the linear stability results. At steady state, the active stress (\(\sim \alpha S \theta\)) and the viscous stress (\(\sim\mu v/W \)) balance
\begin{equation}
    v \sim \alpha S \theta \frac{W}{\mu},
    \label{eqn stress balance}
\end{equation}
where \(\theta\) is the characteristic angle the director field. Upon spontaneous flow transition, we can consider $\theta$ to have the form:
\begin{equation}
    \theta \sim \sqrt{\frac{W-W_c}{W_c}}.
    \label{eqn theta scaling}
\end{equation}
Thus, the angle scales with the reduced width, which acts as a non-dimensional function of the channel width akin to a second order phase transition. Since \(W\approx W_c\) near the transition, combining Eq.~\ref{eqn crtitcal activity flat}$-$\ref{eqn theta scaling} gives the velocity of the active nematic fluid
\begin{equation}
    \frac{v}{\alpha^{1/2}} \sim \sqrt{\frac{W-W_c}{W_c}},
\end{equation}
explaining the collapse of the simulation results shown in Fig.~\ref{fgr:fig2}(d). Therefore, the AN-MPCD simulations of fluid confined within a flat-walled channel are in agreement with the investigations of the existing literature.

\subsection{Coherent flows in flat-walled channel}
\label{3.2}
\begin{figure*}[t!p!]
    \centering
    \includegraphics[scale = 0.53]{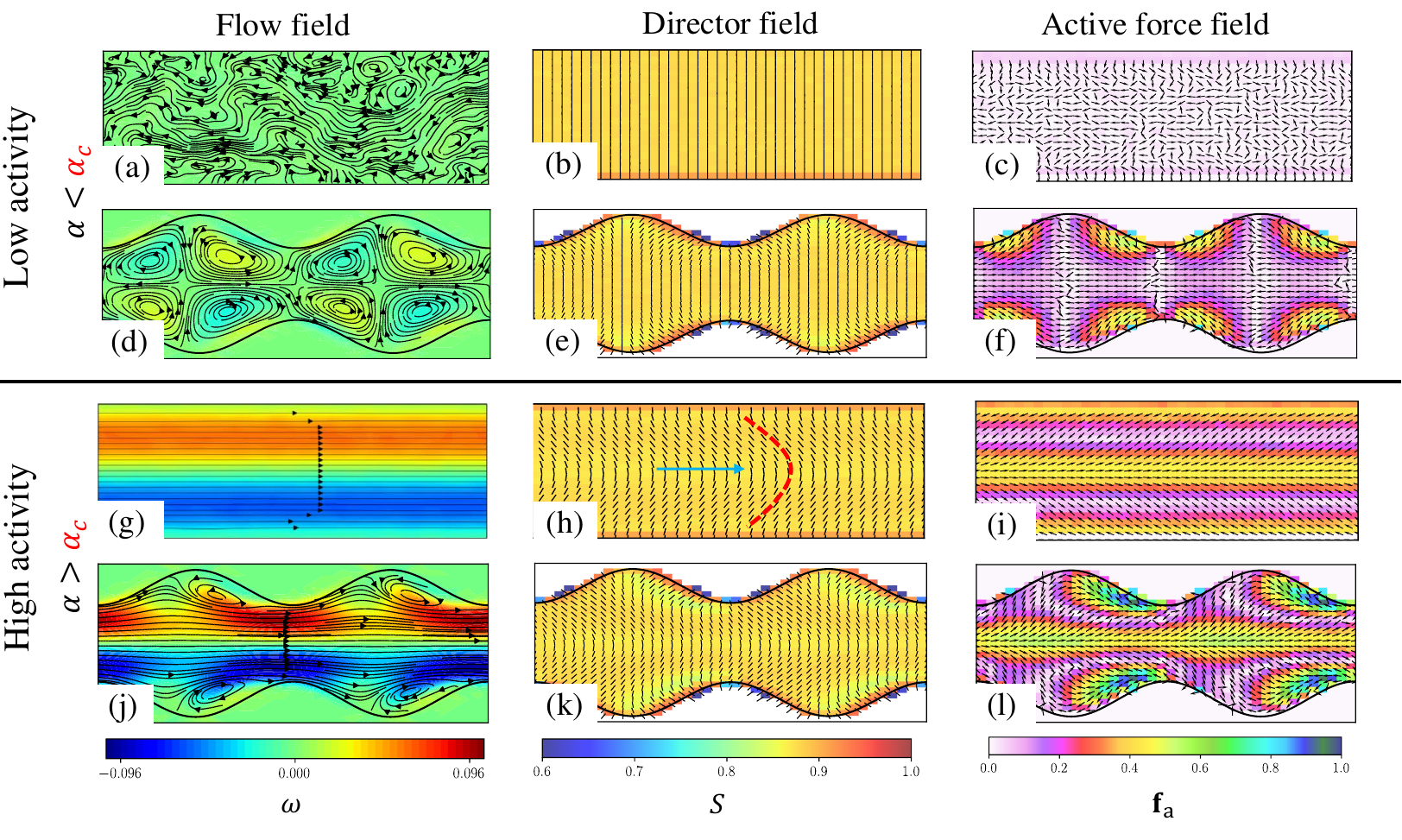}
    \caption{Snapshots of the flow field (left column), director field (middle column) and active forcing (right column) in flat and corrugated channels at lower ($\alpha < \alpha_c$) and higher ($\alpha > \alpha_c$) activity. In the left column, the flow field is shown, which is represented by streamlines superimposed on the vorticity of the fluid. The colorbar is shown at the bottom  with blue and red representing clockwise and anticlockwise rotation respectively. In the middle column, the director field is shown by black dashed line color shaded by scalar order parameter. In the right column, the active force is illustrated by black arrows and are color coded by the magnitude of the force. Figures (a)-(f) correspond to  low activity and figures (g)-(l) correspond to higher activity at which coherent flows prevail. All channels have a width of $W=20$ and corrugated channels have amplitude of $A=3$ and wavelength $\Lambda=20$. For the flat-walled channel $\alpha_c = 0.015$; the cases of low and high activity correspond to $\alpha = 0.009$ and $\alpha = 0.017$ respectively. For the corrugated channel $\alpha_c = 0.019$; the cases of low and high activity correspond to $\alpha = 0.017$ and $\alpha = 0.023$ respectively.}
    \label{fgr:wavyvsflat}
\end{figure*}

When confined, active nematic fluid undergoes a spontaneous flow transition, a coherent flow is usually observed immediately above the transition point \cite{Santhan2D,Santhan3D,Abhik,helical}. Coherent flow is characterized by velocity fields that are dominant along the length of channel \(v_x \approx |\mathbf{v}|\). We now discuss the underlying physical mechanism associated with flow transition to a coherent flow in a flat-walled channel (Fig.~\ref{fgr:wavyvsflat}). The flow field is plotted by superimposing streamlines on fluid vorticity, \(\omega = \left(\nabla \times \mathbf{v} \right)_z\), where the colors represent the clockwise (blue) and anticlockwise (red) rotation of the fluid elements. The local director field \(\mathbf{n}\) is shown by black dashed line, color shaded by the scalar order parameter, \(S\). Active force, \(\mathbf{f}_a\), produced by the nematic fluid is calculated as \(\mathbf{f}_a \propto \nabla \cdot \mathbf{Q}\) and is shown with black arrows depicting the direction and color shading depicting magnitude of the force produced.

For a lower activity (\(\alpha<\alpha_c\)), in a flat-walled channel a no-flow state is observed in the flow domain (Fig.~\ref{fgr:wavyvsflat}(a)), with director field aligned in the nematic phase (Fig.~\ref{fgr:wavyvsflat}(b)). The active forcing produced by the fluid is insignificant (Fig.~\ref{fgr:wavyvsflat}(c)). However, at a higher activity (\(\alpha>\alpha_c\)), a bend-like deformation of the director field is observed (Fig.~\ref{fgr:wavyvsflat}(h)) in which, the thick dashed line is a visual representation of the bend deformation and the arrow represents the direction of fluid flow. The active force produced by the fluid corresponding to the bend deformation is maximum at the center of the channel and oriented along the \(x\) axis (Fig.~\ref{fgr:wavyvsflat}(i)). Thus, the active force drives the flow along the length of the channel producing coherent flows (Fig.~\ref{fgr:wavyvsflat}(g)).

\subsection{Coherent flows in a corrugated channel} 
\label{3.3}
In the case of a corrugated channel, when the activity \(\alpha\) is lower than a critical value \(\alpha_c\), swirling flows, constituted by closed streamlines are observed. This is illustrated in Fig.~\ref{fgr:wavyvsflat}(d) for an amplitude of $A=3$ and wavelength $\Lambda = 20$, demonstrating a significant difference from the no-flow state observed in a flat-walled channel when $\alpha < \alpha_c$.

\begin{figure}[b!h!]
 \centering
 \includegraphics[width=8cm]{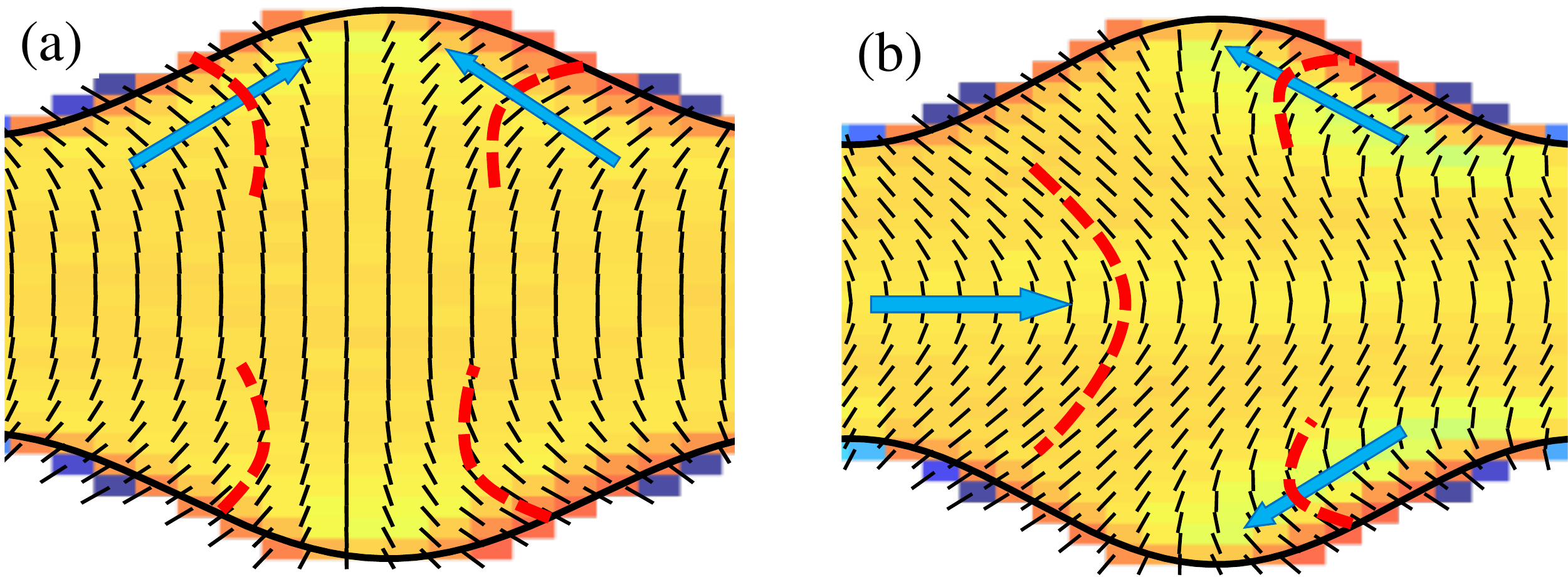}
 \caption{An enlarged view of the director field in the corrugated channel (a) below critical value $\alpha<\alpha_c$ from Fig.~\ref{fgr:wavyvsflat}(e) and (b)  above the critical value $\alpha > \alpha_c$ from Fig.~\ref{fgr:wavyvsflat}(k). The red dashed lines represent the bend like deformation and the blue arrows point in the direction of the active forcing and subsequent fluid flow. }
 \label{fgr:bend_near_walls}
\end{figure}

The origin of swirling flows can be explained by observing the director field (Fig.~\ref{fgr:wavyvsflat}(e)). Due to strong homeotropic anchoring boundary condition, the director field aligns perpendicular to the wall. However, due to the reflection symmetry about the channel centerline, the director field within the bulk of the channel remains oriented perpendicular to the centerline. Therefore, the nematic elasticity induces a bend-like deformation close to the channel wall which is visually depicted by thick dashed lines in Fig.~\ref{fgr:bend_near_walls}(a). As a consequence of this bend, the fluid experiences an active force near the walls (Fig.~\ref{fgr:wavyvsflat}(f)), where the magnitude of force is maximum at the bends. This active force drives the flow towards the centre of the corrugation, as indicated by the arrows in  Fig.~\ref{fgr:bend_near_walls}(a). Since the corrugation is symmetric with respect to the \(y-\)axis, two bends are observed near the wall, facing each other. As a consequence, the flow converges at the center of the corrugations and is subsequently driven towards the bulk of the channel. This dynamics leads to the formation of counter-rotating swirls. Due to the symmetry of the channel, a pair of bends are observed at both the top and bottom walls, resulting in four swirls within the fluid (Fig.~\ref{fgr:wavyvsflat}(d)). The swirling flows prior to spontaneous flow transitions have been analyzed by \citet{Joanny}. Calculations in the limit $A/\Lambda \ll 1$ and $ W \ll W_c \ll \Lambda$ showed that the velocity of the swirls depends on the geometry of the channel as $\sim\frac{A}{\Lambda^4}$. 

At higher activities (\(\alpha>\alpha_c\)) in the corrugated channels, coherent fluid flow is obtained alongside the swirls near the walls (Fig.~\ref{fgr:wavyvsflat}(j)). From Fig.~\ref{fgr:wavyvsflat}(k) we observe that a bend like deformation can be noticed in the bulk of the channel (Fig.~\ref{fgr:bend_near_walls}(b)), which results in coherent flow in the channel. This is in contrast to lower activities (Fig.~\ref{fgr:bend_near_walls}(a)). The combination of strong anchoring boundary condition and nematic elasticity induces a bend-like deformation in the director field close to the channel wall with opposite polarity to the bend in the bulk (Fig.~\ref{fgr:bend_near_walls}(b)). The bend in the director field close to the wall generates an active force, driving the fluid towards the centre of undulation and then towards the bulk of fluid resulting in swirling flows. From Fig.~\ref{fgr:wavyvsflat}(l), it is evident that the active force corresponding to the bend in the director field close to the wall is pronounced compared to the bulk of the channel. It is these regions of high bend and strong active forcing that drive and sustain the swirling flow alongside the bulk coherent flow. 
The swirls generated remain trapped inside the corrugations and coexist along with coherent fluid flow along the center of the channel (Fig.~\ref{fgr:wavyvsflat}(j)). 
\begin{figure*}[t!]
    \centering
    \includegraphics[width =\textwidth]{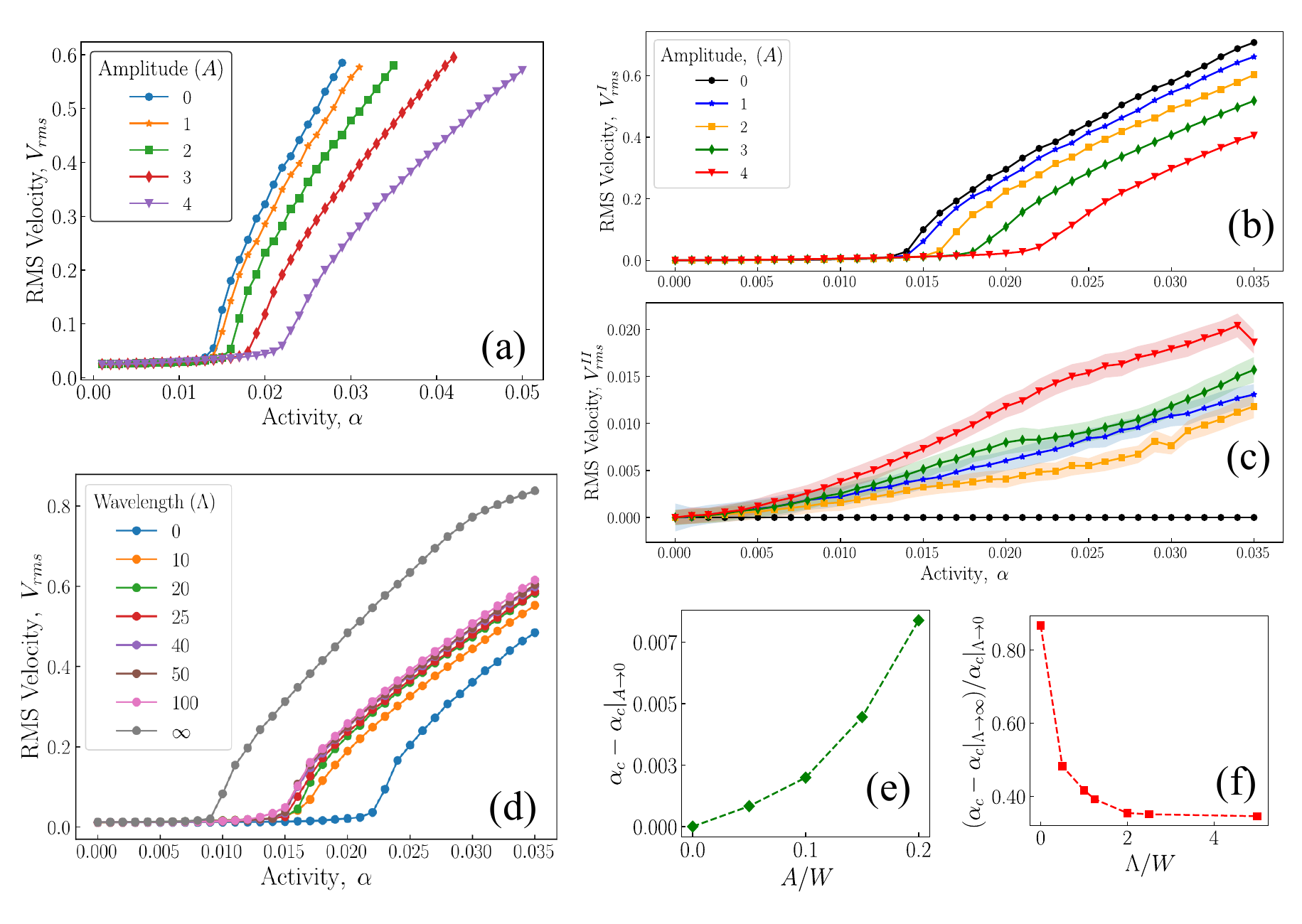}
    \caption{Effect of channel geometry on spontaneous flow transition to a coherent flow state in corrugated channels. Variation in $V_{rms}$ with respect to activity for corrugations of (a) different amplitudes, $A$ but at fixed wavelength $\Lambda =20$ and (d) different wavelengths $\Lambda$ but fixed amplitude $A = 2$. In both cases, the mean width is fixed at $W = 20$. Channel length is chosen as $L = 100$ in (a) and $L = 200$ in (d). Figures (b) and (c) illustrate the variation in $V_{rms}$ as a function of $\alpha$ for fluid within the bulk (region I) and corrugations (region II) of the channel, respectively. Figures (e) and (f) depict the change in critical activity in the corrugated channel as a function of amplitude and wavelength normalised by the mean width $W$ of the channel, respectively.}
    \label{fgr:fig5}
\end{figure*}
It is interesting to note that in a flat-walled channel, for $\alpha>\alpha_c$, fluid at the centerline of the channel experiences a large active force (Fig.~\ref{fgr:wavyvsflat}(i)). In this case, although the active force near the walls is of comparable strength but opposite polarity, it is the active force along the centerline of the channel that dictates the direction of the coherent flow. Similarly, in the corrugated channel the active force is large around the centerline and drives the coherent flow along the length of the channel (Fig.~\ref{fgr:wavyvsflat}(l)). However, the active force experienced by the fluid close to the walls has a higher intensity but with an opposite polarity. 

The existence of coherent flows with swirls align with the experimental observations reported in \citet{Dogic}. When the active nematic fluid (suspension of microtubule and motor protein mixture) is confined in a ratchet like geometry  coherent flows were  produced but were accompanied by  swirls trapped in the corrugations. This observation also emphasizes that the channel geometry can influence the flow dynamics and behavior of an active nematic fluid confined within it.  

\subsection{Effect of changing the amplitude and wavelength of the corrugations}
\label{3.4}
In a flat-walled channel, the active nematic fluid undergoes a flow transition from a state of no-flow to a coherent flow. In contrast, corrugated channels exhibit counter rotating swirls prior to the flow transition. Therefore, we now investigate the effects of change in the corrugation geometry on the spontaneous flow transition of active nematic fluid to a coherent flow state. This is achieved by varying the amplitude, $A$ and wavelength, $\Lambda$ of the corrugations of the channel. 

The results of change in amplitude are shown in Fig.~\ref{fgr:fig5}(a), where root-mean-squared velocity, $V_{rms}$ of the fluid is plotted as a function of activity $\alpha$ for various amplitudes. The active nematic fluid undergoes a flow transition at all amplitudes. However, above the critical activity, the velocity of the fluid decreases with increase in amplitude. Further, it is also observed that larger amplitudes shift the critical activity at higher values. Note that all simulations have been carried out at a fixed mean width $W$ and therefore, increase in the amplitude $A$ reduces the minimum width, $W_{min}$ of the channel. Therefore, the active fluid encounters a stronger geometric constriction with increase in the amplitude of the corrugation, and thus causing the shift of the flow transition to higher activities. 

To determine the critical activity of the active nematic fluid in the corrugated channel the corrugated channel is split into two domains: region I represents the center of the channel, $|y| < (W_{min}/2)$ and region II  represents the fluid trapped in the corrugations $(W_{min}/2) < |y| < (W/2 + A)$ (see the schematic in Fig.~\ref{fig:beta_figs}(a)). The root-mean-squared velocity of the fluid calculated in the central region I $V^I_{rms}$ and in region II within the corrugations, $V^{II}_{rms}$ are shown as a function activity in Fig.~\ref{fgr:fig5}(b) and~\ref{fgr:fig5}(c) respectively. As in a flat channel, $V^I_{rms}$ remains small below the critical activity but increases rapidly as~$\sim\sqrt{\alpha}$ beyond the flow transition. The critical activity $\alpha_c$ corresponding to the corrugated wall is determined from Fig.~\ref{fgr:fig5}(b). On the other hand, the fluid velocity in the corrugations, $V^{II}_{rms}$, continues to gradually increase with increasing activity without a discernible indication of the spontaneous flow transition (Fig.~\ref{fgr:fig5}(c)). The velocity of the fluid trapped in the corrugations $V^{II}_{rms}$ increases with increase in the amplitude of the corrugations. A larger amplitude results in a stronger bend in the director field that generates a  stronger flow in the corrugations. The critical activity required for flow transition determined from Fig.~\ref{fgr:fig5}(a) are shown in Fig.~\ref{fgr:fig5}(e) as a function of amplitude $A$ of the corrugations. The increase in critical activity with amplitude appears to show a quadratic dependence.

To study the effect of the wavelength of the corrugations $\Lambda$ on the active nematic fluid flow, we vary $\Lambda$ for a fixed amplitude and mean width of  the channel. Comparing the root-mean-squared velocity $V_{rms}$ of the fluid as a function of activity $\alpha$ for various wavelengths (Fig.~\ref{fgr:fig5}(d)) to $V_{rms}$ as a function of amplitude (Fig.~\ref{fgr:fig5}(a)) demonstrates that the influence of wavelength in the investigated range is weaker. Changing the wavelength of the corrugations affects the velocity of the coherent flow of active nematic fluid with larger wavelengths resulting in larger velocity. Further, increasing the wavelength of the channel reduces $\alpha_c$ the critical activity required for the spontaneous flow transition (Fig.~\ref{fgr:fig5}(f)). The critical activity decreases with wavelength, showing a strong power law decay. 

\begin{figure*}[t!p!]
    \centering
    \includegraphics[width = \textwidth]{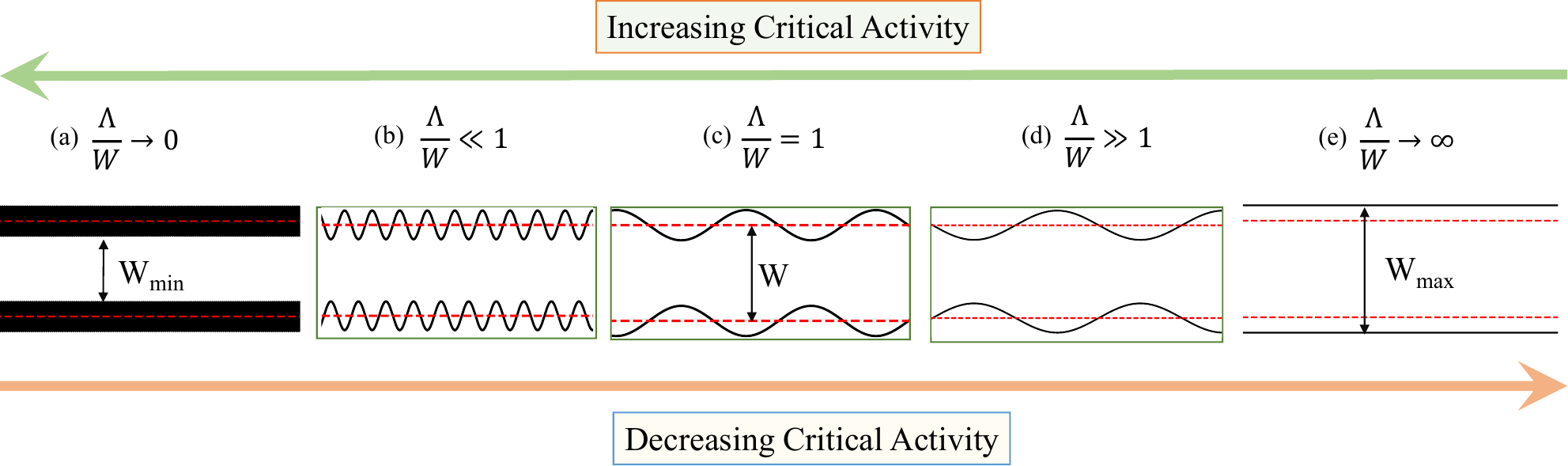}
    \caption{Schematic illustrating the corrugated channel of mean width $W$ as an intermediate construction between a flat-walled channel of width $W_{min} = W - 2A$ and one of larger width $W_{max} = W + 2A$. Corrugated channel approaches flat-walled channel of width $W_{min}$ and $W_{max}$ when $\Lambda/W \to 0$ and $\Lambda/W \to \infty$ respectively. The green arrow at the top shows the increase in critical activity for a corrugated channel of wavelength $\Lambda$ when compared to that of a channel of width $W_{max}$. The yellow arrow at the bottom shows the decrease in critical activity for a corrugated channel when compared to that of a flat-walled channel of width $W_{min}$.}
    \label{fgr:fig6}
\end{figure*}
A corrugated channel can be perceived as a geometry in between two flat-walled channels of different width, as schematically illustrated in Fig.~\ref{fgr:fig6}. The two limiting cases to $\Lambda\to0$ and $\Lambda\to\infty$ both approach the geometry of flat-walled channels with $W \approx W_{min}$ and  $W \approx W_{max}$ respectively  (see Fig.~\ref{fgr:fig6}). In the simulations analyzed above $\Lambda/W \approx 1$ as indicated by the illustration in panel (c) in Fig.~\ref{fgr:fig6}. Reducing  $\Lambda/W$ leads to a denser packing of corrugations (panel (b)). The limiting case $\Lambda /W\to0$ (panel (a)) corresponds to a flat-walled channel of width $W_{min}$ . On the other hand, an increase in $\Lambda/W$ leads to a channel geometry shown in panel (d) with limiting case $\Lambda/W \to\infty$ (panel (e)), that again corresponds to a flat-walled channel but now of width $W_{max}$. Therefore, to first order any corrugated channel can be thought of as a construction intermediate between the limits of two flat-walled channels of width $W_{min}$ and $W_{max}$. As a consequence, the critical activity required for spontaneous flow transition of an active nematic confined in a corrugated channel falls between the critical activity of two flat-walled channels of widths $W_{min}$ and $W_{max}$. This was also evident in Fig.~\ref{fgr:fig5}(b) where the root-mean-squared velocity data was plotted against activity for various wavelengths.

Since critical activity is inversely proportional to the square of channel width, critical activity required for spontaneous flow transition in a flat-walled channel of width $W_{min}$ is larger than that in channel of width $W_{max}$. Therefore, the critical activity required for the spontaneous flow transition in intermediate configurations, will be smaller than that of a flat-walled channel of width $W_{min}$ but greater than that of a flat-walled channel of width $W_{max}$. 


Yet another way to perceive the coherent flows in corrugated channels when wavelength is increased from that corresponding to $W_{min}$ to $W_{max}$ is the reduction in the number of geometrical constrictions or correspondingly the increase in the volume occupied by the fluid between the crests, where swirls are present. Hence the reduction in critical activity with increase in wavelength can also be thought of as providing an effective slip to the fluid in the bulk.

\begin{figure*}[t!]
 \centering
\includegraphics[width =\textwidth]{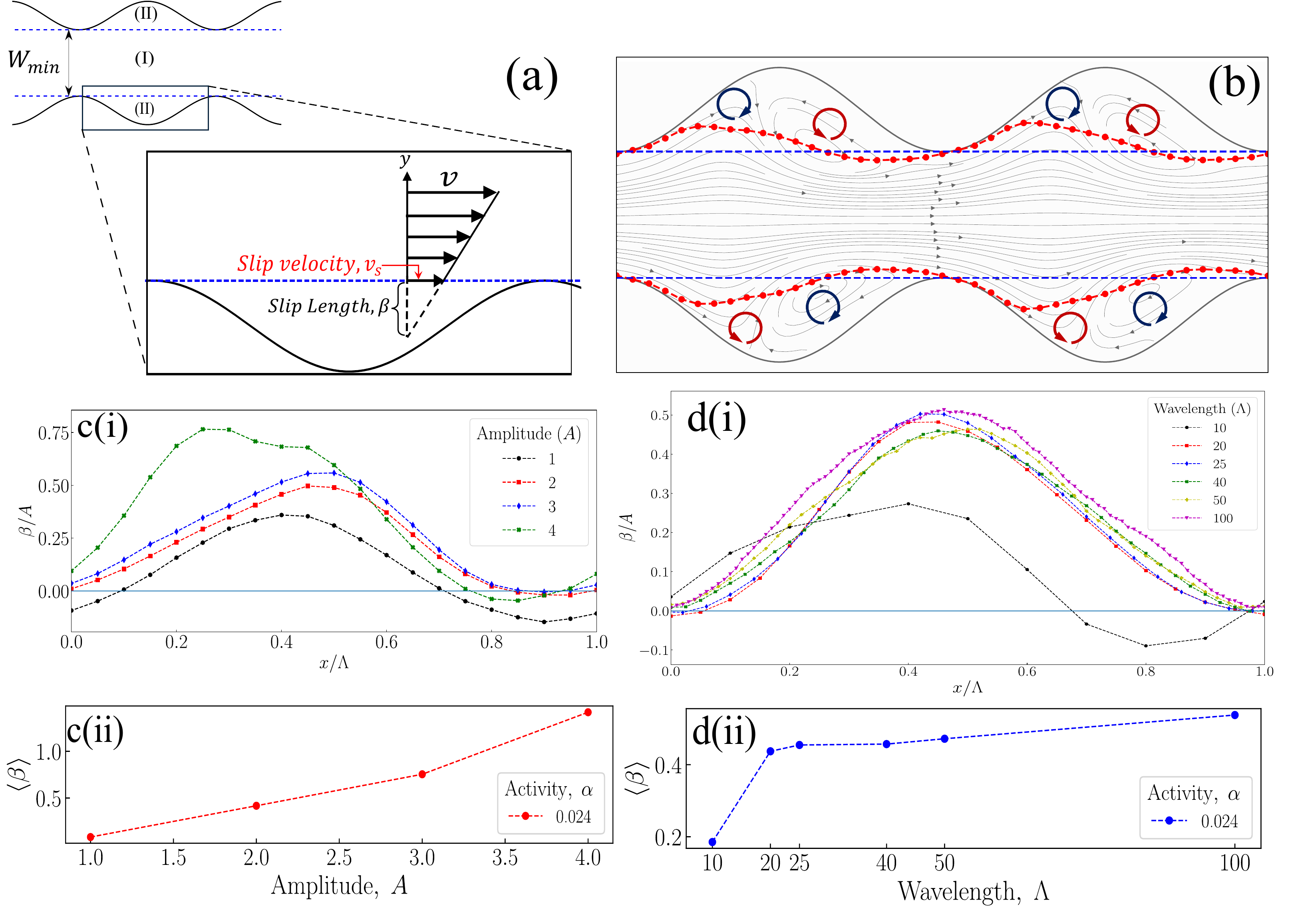}
 \caption{(a) Schematic illustrating the demarcation of the active nematic fluid domain into two regions and the definition of slip length. Region I when $|y| < (W_{min}/2)$ where the bulk of the fluid is located and region II when $(W_{min}/2) < |y| < (W_{max}/2)$ where the swirls are present. Fluid in region I experiences a slip velocity due to the presence of fluid in region II. (b) Slip length calculated at the boundary of region I and II. The boundary at $y = \pm W_{min}/2$ is indicated by blue, dashed lines while the boundary accounting for slip length \textit{i.e.,} $y(x) = \pm W_{min}/2 \pm \beta(x)$ is shown by the red, continuous line connecting the symbols. The circular arrows indicate the sense of rotation of the vorticity associated with the flow in the corrugations. (c) Variation in slip length $\beta(x)$ for various channel wall amplitudes with $\Lambda = 20$ and (ii) average slip length $\langle \beta \rangle$ for corresponding amplitudes. (d) Variation in slip length $\beta(x)$ for various wavelengths with $A =2$ (x-axis is normalised by $\Lambda$) and $\langle\beta\rangle$ as a function of wavelength.}
\label{fig:beta_figs}
\end{figure*}

\subsection{Effect of swirls on the coherent flow of active nematics in corrugated channels}
\label{3.5}
We have seen that coherent flows in the center of a corrugated channel are accompanied by swirls trapped in the corrugations (Fig.~\ref{fgr:wavyvsflat}(j)). The presence of swirls in region II (Fig.~\ref{fig:beta_figs}(a)) uniquely differentiates the coherent flows in corrugated channels compared to flat-walled channels. To elucidate the role of swirls in supporting coherent flows, we interpret the swirls as a mechanism to provide a slip velocity to coherent flows in the center of the channel (region I; Fig.~\ref{fig:beta_figs}(a)). 

We model the effect of swirling fluid in the corrugations (region II) on the flow in the central region (in region I) as providing a slip at the interface of two regions, $y = \pm W_{min}/2$. This model is schematically illustrated in Fig.~\ref{fig:beta_figs}(a). The extent of slip is characterised by a slip velocity $v_s$ and corresponding slip length $\beta$. The slip length corresponds to a fictitious distance at which the fluid velocity in region II decays to zero if the velocity gradient very close to the boundary in region I is linearly and extrapolated as
\begin{align}
    v_s(x) = \beta \left.\frac{dv}{dy}\right|_{y = \pm W_{min}} .
    \label{slip bc}
\end{align}
The above equation assumes that viscous stress dominates the interface between region I and region II and hence utilises Newtonian constitutive relation in evaluating the slip length $\beta$. Due to the variation in the geometry along the channel length, slip length is not a constant. The function $\beta (x)$ accounts for the effect of the fluid that is present in the corrugations on the coherent flow in the center of the channel.

An example of the slip length $\beta (x)$ extracted from Eq.~\ref{slip bc} for a coherent flow state in a corrugated channel is shown in Fig.~\ref{fig:beta_figs}(b). The horizontal dashed blue line demarcates region I from region II. The dashed red lines connecting the symbols indicate the slip length $\beta(x)$ from the boundary $y = \pm W_{min}$. As expected the slip length goes to zero at the constriction, where the interface separating region I and II coincides with the rigid boundary. The slip length is a periodic function but not mirror symmetric with respect to its zero value due to the asymmetry of the swirls in region II (Fig.~\ref{fgr:wavyvsflat}(j)). The slip length is positive on the left and negative on right hand side of the corrugation when the coherent flow is from left to right, as shown in Fig.~\ref{fig:beta_figs}(b). The positive slip length indicates a support in maintaining the coherent flow. The negative slip length indicates the resistance offered by the swirls towards the coherent flow on the right hand side of the corrugation, which occurs due to the presence of the counter clockwise rotating swirls in this case. However, as evident from the Fig.~\ref{fig:beta_figs}(b), the extend of positive slip on the left side is more than that on the right side, indicating that the fluid in region I experiences a net positive slip due to the presence of fluid in region II. Thus, the overall effect of active nematic fluid entrapped in the corrugations is to endow a positive slip to the coherent flow in the center of the channel.

Since, the amount of fluid contained in the corrugations varies with the amplitude and wavelength which changes the behaviour of swirls in corrugations, therefore, the slip length $\beta(x)/A$ is plotted for various amplitudes and wavelengths of the corrugations in \ref{fig:beta_figs}(c)-(d). In Fig.~\ref{fig:beta_figs}c-d(i), the abscissa is normalised with $\Lambda$ for comparison between various cases. It is clear that the slip length increases with amplitude. 

The variation with wavelength is more intricate (Fig.~\ref{fig:beta_figs}d(i)). To better understand how the effective slip length calculated from Eq.~\ref{slip bc} varies with wavelength, we average the slip length over an integer number of wave, $\langle\beta\rangle=\int_0^\Lambda\beta(x)dx/\Lambda$. The average slip length is plotted separately as a function of amplitude and wavelength in \ref{fig:beta_figs}c-d(ii). Increasing the amplitude increases the average slip length monotonically. However, increasing the wavelength seems to show two distinct behaviours: when the wavelength is small ($\Lambda/W < 1$), slip length increases with increase in wavelength but it becomes independent of wavelength for large wavelength ($\Lambda/W>1$). 

To understand the implications of these variations we analyze the flow transition in corrugated channels through an analytical approach. Since the fluid in the center of the channel (region I) exhibits a flow transition similar to that in a flat-walled channel, the role of the slip on the spontaneous flow transition may be analyzed using a linear stability analysis (see Appendix C for detailed calculations). 
\begin{figure*}[tp]
 \centering
    \includegraphics[width = \textwidth]{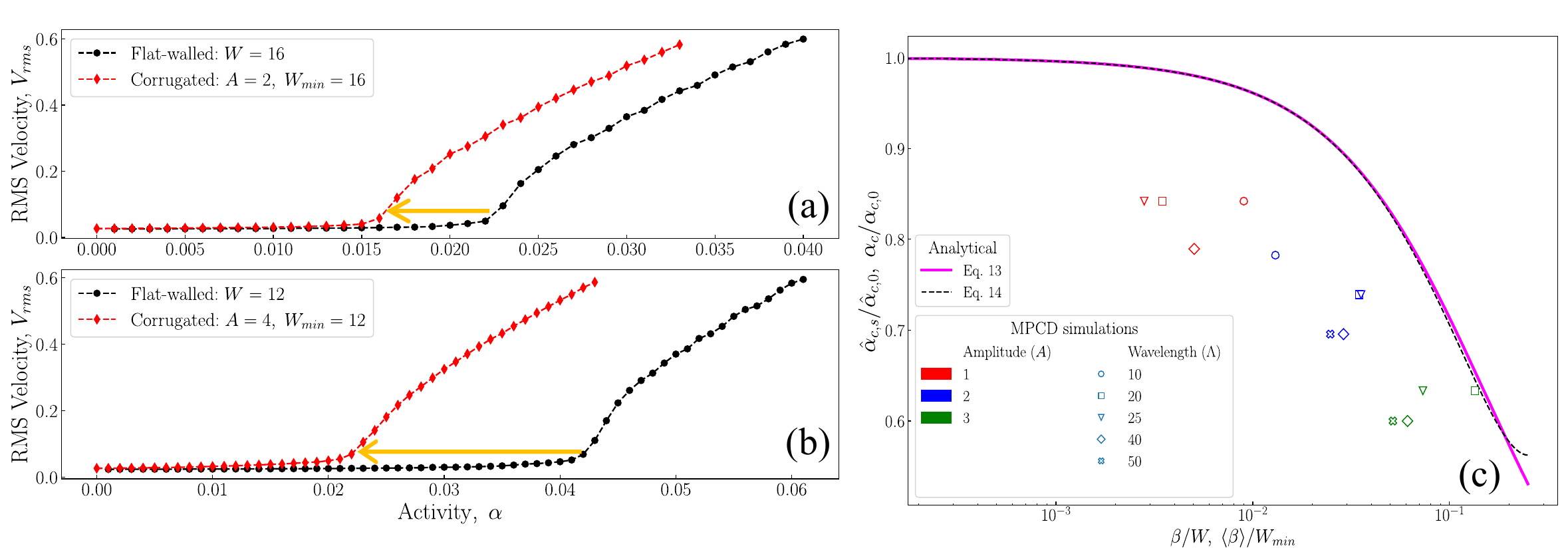}
 \caption{Reduction in the critical activity for spontaneous flow transition in a corrugated channel of minimum width $W_{min}$ compared to a flat wall channel of width $W = W_{min}$. $V_{rms}$ is plotted as a function of $\alpha$ for (a) a corrugated channel $W_{min} = 16,~A = 2$ in comparison with a flat-walled channel $W = 16$ and (b) a corrugated channel $W_{min} = 12,~A = 4$ in comparison with flat-walled channel $W = 12$. (c) Results from the linear stability analysis - the critical activity in a flat-walled channel with slip boundary conditions, $\alpha_{c,s}$, normalised with that in a channel with no-slip boundary conditions, $\alpha_{c,0}$ is plotted against the slip length as a continuous line (full solution given Eq.~\ref{eqn:Teqn}). The approximate solution given by Eq.~\ref{approximate sol} is shown as the dashed line. Critical activity for the corrugated channel, $\alpha_c$ normalised with that of a flat-walled channel $\alpha_{c,0}$, obtained from AN-MPCD simulations, is plotted against the normalised average slip length, $\langle\beta\rangle /W_{min}$ is shown as markers. }
 \label{fgr:fig10}
\end{figure*}
We consider an active nematic fluid contained in a flat-walled channel of width $W$. The channel walls are characterised by a slip length $\beta$, resulting in a slip velocity at the channel boundary. A quasi 1D channel geometry is assumed, \textit{i.e.,} flow occurs only along the channel length but neither the flow nor director field exhibit any variations along the channel length. As described above we assume that viscous stresses dominate the boundary and hence Eq.~\ref{slip bc} based on Newtonian constitutive relation is used as a boundary condition on the channel walls. Following \cite{Giomi_2012, R.Voituriez_2005} the linearised governing equations for the director field $\theta(y)$ and the stream-wise velocity field $v_x(y)$ are
\begin{align}
        \partial_t \theta &= \gamma^{-1} K \partial_y^2 \theta - \frac{\partial_y v_x}{4}(2-\Hat{\lambda} \cos 2\theta)
    \label{theta final1}\\
        \rho \partial_t v_x &= \partial_y \Bigl( \mu \partial_y v_x +\Hat{\alpha} S \sin 2\theta \Bigr). 
    \label{v final1}
\end{align}
Subjected to the slip boundary conditions, $v_x (y=0~\text{and}~y=W) = \beta \frac{dv}{dy}$ and strong homeotropic anchoring $\theta(y=0~\text{and}~y=W) = \frac{\pi}{2}$, we seek stationary solutions of Eqs.~\ref{theta final1}-\ref{v final1}. The  condition for the existence of flow for a nonzero activity is found to be,
\begin{equation}
    1 - \cos TW + \beta T \sin TW =0,
    \label{eqn:Teqn}
\end{equation}
where $T = \sqrt{\frac{\Hat{\alpha}(2+\Hat{\lambda})S}{2\mu \gamma^{-1} K}}$ (see Appendix~\ref{Appendix C}) is an inverse length scale. Together, $TW$ represents a dimensionless activity number that represents the competition between confinement and active nematicity. The solution of Eq.~\ref{eqn:Teqn} prescribes the critical activity $\Hat{\alpha}_{c,s}$ required for spontaneous flow transition in a flat-walled channel endowed with slip boundary conditions. An explicit expression for critical activity cannot be written down due to the non-linearity of Eq.~\ref{eqn:Teqn}, and must be found numerically. However, for small slip lengths ($\beta/W \to 0$), Eq.~\ref{eqn:Teqn} can be further simplified to obtain an explicit expression for critical activity
\begin{equation}
    \Hat{\alpha}_{c,s} = \Hat{\alpha}_{c,0} \left[1-2\left(\frac{\beta}{W}\right) +4 \left(\frac{\beta}{W}\right)^2\right]^2,
    \label{approximate sol}
\end{equation}
where $\Hat{\alpha}_{c,0}$ is the critical activity required for spontaneous flow transition in a flat-walled channel with no-slip boundary conditions (Eq.~\ref{eqn crtitcal activity flat}). Clearly, Eq.~\ref{approximate sol} recovers the correct limit of $\Hat{\alpha}_{c,s}\to \Hat{\alpha}_{c,0}$ when the slip length, $\beta = 0$. Further, the critical activity is reduced in a channel with slip boundary conditions compared to that with no-slip boundary conditions, and the reduction is proportional to square of the slip length at leading order.

We now compare the results of the linear stability analysis with that from AN-MPCD simulations in Fig.~\ref{fgr:fig10}. Firstly, the root-mean-square velocity $V_{rms}$ is plotted against activity for a corrugated channel of $W_{min} = 16, A = 2$ in comparison with a flat-walled channel of width, $W = 16$ (Fig.~\ref{fgr:fig10}(a)) and in Fig.~\ref{fgr:fig10} (b) $V_{rms}$ is plotted against activity for a corrugated channel of $W_{min} = 12, A = 4$ in comparison with that in a flat-walled channel of width, $W = 12$. In each figure the fluid in the corrugated channel undergoes spontaneous flow transition at a lower activity compared to that of a flat-walled channel. This reduction is indicated in the figure using yellow arrows. The reduction in the value of critical activity for the corrugated channel can be attributed to the slip velocity provided by the fluid trapped in the corrugations, consistent with the predictions of the linear stability analysis. The fluid in the center of the channel (region I) is separated from the no-slip walls because of the presence of region II. On the other hand, in a flat-walled channel, the fluid is always in contact with the channel wall and the no-slip boundary condition on the channel walls mandates a higher activity for a spontaneous flow transition.

We now compare the prediction of the linear stability analysis, Eq.~\ref{eqn:Teqn}, quantitatively with the results from AN-MPCD simulations. The critical activity $\Hat{\alpha}_{c,s}$ in a flat-walled channel endowed with a slip velocity is plotted as a function slip length $\beta$ (Fig.~\ref{fgr:fig10}(c)). In this plot, the critical activity is normalised by $\Hat{\alpha}_{c,0}$, the critical activity in a rigid channel that imposes no-slip boundary conditions (Eq.~\ref{eqn crtitcal activity flat}). The slip length is normalised with the channel width $W$. As expected, both the full solution (Eq.~\ref{eqn:Teqn}) and the approximate solution (Eq.~\ref{approximate sol}) of the critical activity decrease with increased slip length. The AN-MPCD data confirms that the critical activity decreases with slip length. The significant match between the results from the linear stability analysis and the AN-MPCD simulations suggests that the effect of active swirling flows that are generated in the corrugations due to the distortions in the director field is equivalent to an effective slip for the active coherent flow that develops in the center of the corrugated channel.

\section{Conclusions}
\label{sec:conclusion}
In this work, we perform numerical simulations using an active nematic multi-particle collision dynamics (AN-MPCD) algorithm to study the spontaneous flow transition and the resulting flow state of an active nematic confined in a corrugated channel. Active nematics confined between two parallel plates is a well studied system\cite{ucmerced,Joanny,D.Marenduzzo}. At low activities, the active nematic fluid is stationary with nematic elasticity dominating the system. At sufficiently high activity, the nematic fluid undergoes a spontaneous flow transition. The resulting velocity of the fluid is proportional to the square root of activity, while the critical activity is inversely proportional to square of the channel width. In this work, we first demonstrated these scalings for an AN-MPCD fluid.

We extended channel-confined active fluids to study the behaviour of active nematics confined in corrugated channels. In this study, the wavy walls on the top and bottom of the corrugated channel are out-of-phase, but  qualitatively similar conclusions are reached in a snaking channel in which the bottom and top walls are in-phase (Appendix~\ref{Appendix D}). We find that the distinguishing feature of the spontaneous flow transition in corrugated channels is from a weak flow state to a strong flow state. This is in contrast to a flat-walled channel where the transition is from a no-flow state to a flowing state. The weak flows before the transition originate from distortions in the director field arising from the preferred anchoring of the active nematic to the curved boundaries. The critical activity required for flow depends upon the channel geometry. Increasing the amplitude of the corrugations while keeping the mean width of the channel fixed increases the constrictions in the channel and thus increases the critical activity required for flow transition. On the other hand, changing the wavelength of the corrugations can be understood by regarding the corrugated channel as intermediate between two flat-walled channels of different widths, $W_{min}$ and $W_{max}$. The critical activity for the corrugated channel is smaller than a flat-walled channel of width $W_{min}$ and higher than a flat-walled channel of width $W_{max}$. The current study limited itself to the transition to unidirectional flow. However, it is well-known that bidirectional flow \cite{Duclos2018,Li2021,Pratley2024} and higher modes frequently occur\cite{doostmohammadi2018active,Giomi_2012,D.Marenduzzo} in flat channels. Future works, should consider how undulations, slip-length or roughness effect the stability of bidirectional flow states.

At activities above the flow transition, the boundary-induced swirling flows are confined to the corrugations. These swirls can be contrasted with Newtonian flows in corrugated micro-channels: (i) Newtonian fluids at low Reynolds number do not generate swirls in a symmetric corrugated channels due to reversibility\cite{Leal_2007} and (ii) swirls generated by a Newtonian fluid at high Reynolds number are driven by the flow in the center of the channel\cite{tatsuo,tolentino}. In contrast the swirls observed in the active nematics in corrugations are generated by the active stress arising from the distortions in the director field. Our analysis shows that the effect of these swirls can be modelled as an effective slip to the coherent fluid flow that occurs in the bulk. Using this principle, we performed a linear stability analysis to determine the effect of slip mechanism on the spontaneous flow transition, and matched the results from the simulations of corrugated channels with predictions from the stability analysis. Our study reveals the physical mechanisms at play when active nematics are confined in corrugated channels and further highlights the significant role of the boundaries in determining the dynamics of confined active systems. 

Finally, our studies also suggest the possible mechanism by which asymmetric notches/teeth on the channel wall can direct coherent flows of channel confined active nematics, as seen in experiments \cite{Dogic}. Active swirls generated inside the notches due to deformations in the director field provide a slip to the coherent flow; and thus, the chirality of the stronger and bigger swirl of the asymmetric notches dictate the direction of the flow in the center of the channel. Future studies should focus on validating this hypothesis and investigating the role of asymmetry of the channel corrugations in dictating the direction of coherent flows.


\section*{Conflicts of interest}
There are no conflicts to declare.

\section*{Appendix}
\appendix

\section{Nematic Multi-Particle Collision Operator}
\label{Appendix A}
    The nematic collision operator $\boldsymbol{\Xi}_i^N$ is a version of the Anderson-thermostatted collision operator~\cite{Noguchi2007,Tyler2015}
    \begin{equation}
        \boldsymbol{\Xi}_{i}^N = \boldsymbol{\xi}_j - \left \langle \boldsymbol{\xi}_j \right \rangle  + \left (\boldsymbol{I}^{-1} \cdot \left (\delta \mathbf{L}_{vel} + \delta \mathbf{\mathcal{L}}_{ori} \right)\right) \times \boldsymbol{r}_i' ,
        \label{anderson operator}
    \end{equation}
    where $\boldsymbol{\xi}_{i}$ is a random velocity drawn from the Maxwell-Boltzmann distribution for a thermal energy $k_\text{B} T$ and $\left \langle \boldsymbol{\xi}_j \right \rangle$ is the average of the random velocities generated for all the particles within each cell. Subtracting the average velocities ensures conservation momentum when all MPCD particles have the same mass. The instantaneous moment of inertia for a cell is given as  $\mathbf{I} = \sum_j^{\Tilde{N}} m_j \left(\mathbf{r}_j'^2 \Hat{\mathbf{1}} - \mathbf{r}_j'\mathbf{r}_j'\right)$ for point particles relative to the cell's center of mass $\mathbf{r^{cm}}$, where $\mathbf{r}_i' = \mathbf{r}_i - \mathbf{r^{cm}}$. The third term in Eq.~\ref{anderson operator} removes spurious angular momentum introduced by collision operation $\delta \mathbf{L}_{vel}  = \sum_j^{\Tilde{N}} m_j\left (\mathbf{r}_j' \times \left(\mathbf{v}_j - \boldsymbol{\xi}_j\right)\right)$ or by the rotation of the nematogens $\delta \mathbf{\mathcal{L}}_{ori}$. 
    Including $\delta \mathbf{\mathcal{L}}_{ori}$ in $\boldsymbol{\Xi}_i^N$ ensures conservation of angular momentum and accounts for the nematic back flow. 
    The rotation of the nematogens arise from the stochastic orientational collision operation (Eq.~\ref{eqn:orientation collision}) and from the response of nematogens to velocity gradients, which is accounted for through Jeffery's equation
    \begin{equation}
        \delta \mathbf{u}_i^J = \delta t \chi \left[\mathbf{u}_i \cdot \boldsymbol{\Omega} + \lambda \left (\mathbf{u}_i \cdot \mathbf{E} - \mathbf{u}_i\mathbf{u}_i\mathbf{u}_i:\mathbf{E}\right)\right] 
    \end{equation}
    where $\lambda$ is a bare tumbling parameter and $\chi$ is the heuristic shear coupling coefficient in a flow with shear rate tensor $\mathbf{E} = [\mathbf{\nabla v} + (\mathbf{\nabla v})]/2$ and vorticity tensor $\boldsymbol{\Omega} = [-\mathbf{\nabla v} - (\mathbf{\nabla v})]/2$. The heuristic shear coupling coefficient $\chi$ tunes the hydrodynamic susceptibility of orientation to velocity gradients. MPCD particles rotated by torques are balanced by applying equal and opposite torque to fluid by a change in angular momentum $\delta \mathbf{\mathcal{L}}_{ori}(t) = -\gamma_R \sum_i^{\Tilde{N}} \mathbf{u}_i(t) \times \dot{\mathbf{u}}_i$, where $\gamma_R$ is the viscous rotation coefficient.

\section{Corrugated boundary implementation}
\label{Appendix B}

In AN-MPCD, boundary conditions comprise two pieces: \textit{Boundary surfaces}, which are sinusoids, and the \textit{boundary rules} which are applied to fluid particles when they violate the boundary surfaces. 
To implement wavy walls we follow \citet{Wamsler2024}

\subsection{Boundary Surfaces}
\label{sctn:surfaces}
The boundaries are represented by surfaces $\mathcal{S}_{b}(\mathbf{r})=0$ expressed in the form  
\begin{equation}
    \mathcal{S}_{b,0}(\mathbf{r}) = \mathbf{A}_{b} \cdot \left( \mathbf{r} - \mathbf{q}_{b} \right)  + B_{b}\cos\left(\frac{2\pi}{\Lambda_b} \frac{ (\mathbf{r}\times \mathbf{A}_{b})\cdot\hat{z} }{ \left|\left|\mathbf{A}_{b}\right|\right|}  \right) ,
    \label{surface eqn}
\end{equation}
where $b$ is the index of the surface. 
The first term is the equation of a plane, where $\mathbf{q}_b$ sets the position of the $b$\textsuperscript{th} boundary and $\mathbf{A_b}$ denotes the plane's normal vector. 
The second term incorporates the corrugations with amplitude $B_{b}$ and wavelength $\Lambda_b$. 
Even though the simulations are in 2D, the cross product and projection on $\hat{z}$ inside the cosine are written for conciseness. 
More complicated functions are required for non-planar, wavy surfaces in 3D~\cite{Wamsler2024}.

When the $i$\textsuperscript{th} MPCD particle resides on the $b$\textsuperscript{th} surface, $\mathcal{S}_{b}(\mathbf{r}_i)=0$. 
 At all other points, $\mathcal{S}_{b}(\mathbf{r}_i)\neq0$. 
 Whenever the MPCD particle is inside the control volume $\mathcal{S}_{b}(\mathbf{r}_i)>0$, but whenever the MPCD particle has impinged on the wall $\mathcal{S}_{b}(\mathbf{r}_i)<0$. 
 If an MPCD impinges on a wall, its ballistic trajectory is ray-traced back to surface and a collision event is initiated. 
 The collision is represented by a set of boundary rules.

\subsection{Boundary rules}
\label{sctn:rules}
The surface boundary defined by Eq.~\ref{surface eqn} exists at $\mathcal{S}_{b}(\mathbf{r}) =0$. Therefore for a particle with position $\mathbf{r}_i$, if  $\mathcal{S}_{b}(\mathbf{r}) \geq 0$, then it is defined as to be into the channel.
Whenever an MPCD particle is outside a boundary, a set of rules defines how the position $\mathbf{r}_i(t)$, velocity $\mathbf{v}_i(t)$ and orientation $\mathbf{u}_i(t)$ are transformed.

The boundary rules are applied in directions that depend on the local surface normal 
\begin{equation}
    \mathbf{n}_b = \left. \frac{\nabla\mathcal{S}_b}{\left|\nabla \mathcal{S}_b\right|} \right|_{\mathbf{r}_i}
\end{equation}
at the collision point. 

Upon crossing the boundary $b$, particle \textit{i}'s position is update to $\mathbf{r}_i \rightarrow \mathbf{r}_i + \mathcal{D}_{b}\mathbf{n}_b$, where
$\mathcal{D}_{b}$ represent a shift in particle's position in the normal direction. Periodic boundary conditions (Section~\ref{sctn:PBC}) set $\mathcal{D}_{b}$ equal to the system size, whereas impermeable walls set $\mathcal{D}_{b}=0$.

The particle velocity upon colliding with the boundary $b$ is updated using parameters $\mathcal{M}_{b,n}$ (normal) and $\mathcal{M}_{b,t}$ (tangential) and is transformed as $\mathbf{v}_i \rightarrow \mathcal{M}_{b,n} \left(\mathbf{n}_b \otimes \mathbf{n}_b\right) \cdot \mathbf{v}_i +  \mathcal{M}_{b,t} \left(\mathbf{1}-\mathbf{n}_b \otimes \mathbf{n}_b\right) \cdot \mathbf{v}_i $. Here, $\left(\mathbf{n}_b \otimes \mathbf{n}_b\right)$ and $\left(\mathbf{1}-\mathbf{n}_b \otimes \mathbf{n}_b\right)$ are normal and tangential projection operators. 
Periodic boundary conditions do not change a particle's velocity and so $\mathcal{M}_{b,n} = \mathcal{M}_{b,t}  = +1$. 
On the other hand, impermeable no-slip boundaries require bounce-back rules with $\mathcal{M}_{b,n} = \mathcal{M}_{b,t} = -1$. 

The orientation of the particle also has multiplicative operators and is transformed to $\mathbf{u}_i \rightarrow \mu_{b,n} \left(\mathbf{n}_b \otimes \mathbf{n}_b\right)_b \cdot \mathbf{u}_i + \mu_{b,t} \left(\mathbf{1}-\mathbf{n}_b \otimes \mathbf{n}_b\right) \cdot \mathbf{u}_i$, which is rescaled back to unit vector. Here, $\mu_{b,n}$ and $\mu_{b,t}$ are normal and tangential orientation parameters used to update particle orientation upon crossing the boundary. 
Free anchoring is implemented through $\mu_{b,n}=\mu_{b,t}=1$ such that the orientation is unchanged. 
Setting $\mu_{b,t}=0$, imposes homeotropic anchoring, while $\mu_{b,n}=0$ results in planar anchoring.
In this study, we consider only homeotropic channel walls ($\mu_{b,t}=0$).

\subsection{Boundary conditions}
\label{sctn:BC}
The system is enclosed by four boundaries. 
Two of these are the left periodic boundary condition (PBC) $b=0$ and the right PBC $b=1$. 
The other two are the impermeable no-slip walls at the bottom $b=2$ and top $b=3$. 

\subsubsection{Periodic boundary condition}
\label{sctn:PBC}
In this study, the PBCs are implemented along the $x$-direction. 
This PBC is achieved by updating the particle location by the length of the simulation domain in direction normal to the boundary through which particle exits. 
Velocity and orientation remain unchanged. 
Therefore, $\mathcal{M}_{b,n} =\mathcal{M}_{b,t} =\mu_{b,n}=\mu_{b,t}=1$, and $\mathcal{D}_{b,n} = L$, where $L\in\left\{100,200\right\}$ is the channel length.
For the PBC on the left of the system, $\mathbf{A}_{0}=\hat{x}$, $\mathbf{q}_{0}=\mathbf{0}$ and $B_0=0$. 
For the PBC on the right, $\mathbf{A}_{1}=-\hat{x}$, $\mathbf{q}_{1}=L\hat{x}$ and $B_1=0$.

\subsubsection{Impermeable, no-slip wavy walls}
\label{sctn:walls}
The bottom wall is defined by the surface parameters $\mathbf{A}_{2}=\hat{y}$ and $\mathbf{q}_{2}=\mathbf{0}$, with variable corrugation parameters $B_2=B$ and $\Lambda_2=\Lambda$. 
Likewise, the bottom wall is defined by $\mathbf{A}_{3}=-\hat{y}$ and $\mathbf{q}_{2}=W\hat{y}$ (for channel width $W=20$), with the same variable corrugation parameters $B_3=B$ and $\Lambda_3=\Lambda$ as the bottom wall.

On these solid walls, the no-slip boundary condition is applied by enforcing the bounce-back rule  $\mathcal{M}_{b,n} =\mathcal{M}_{b,t} = -1$ for $b\in\left\{2,3\right\}$. 
The particle is rewound to the location where it crossed the boundary so that it streams with update velocity for rest of the time step. 
Solid walls do not impose a translational shift, $\mathcal{D}_{b,n} =0$. 
The intersection of boundaries with MPCD cells results in lower particle density as part of the cell is inaccessible to MPCD particles, which in turn lowers the local viscosity~\cite{Noguchi2007} and causes an effective slip. Therefore, to enforce a perfect no-slip boundary condition on solid walls, `phantom' particles are added to ensure correct average density in the cell~\cite{gompper_paper,Bolintineanu2012}.

Here, the walls are chosen to be homeotropic so $\mu_{b,t}=0$ for $b\in\left\{2,3\right\}$ 
However, for the no-slip boundary condition, the boundary rules are insufficient for imposing strong anchoring because the particles that cross the boundary are immediately mixed with particles that have not~\cite{head2024_SM}. 
To ensure strong anchoring, the particles in the cells intersecting the boundary $\mathcal{S}_b$ are all aligned along $\mathbf{n}_b$. 
Since, the intersecting cells are subjected to anchoring condition, this results in generating of strong anchoring~\cite{head2024_SM}.

\section{Linear stability analysis in a flat walled channel with slip boundary conditions}
\label{Appendix C}

Neglecting the variation in the channel geometry and slip length along $x$, we calculate the critical activity for onset of spontaneous flow transition in a flat walled channel endowed with slip boundary conditions.

In addition to equation for conservation of mass, the generalised Navier Stokes equation for the active nematic fluid is given as:
\begin{equation}
    \rho (\partial_t + \mathbf{v}\cdot \nabla \mathbf{v}) = \nabla\cdot \boldsymbol{\sigma}
    \label{convective}
\end{equation}
where, $\rho$ is the density of fluid, $\boldsymbol{\sigma}$ is the total stress tensor and is given as $\boldsymbol{\sigma} = 2\mu \mathbf{E} -P\delta + \boldsymbol{\sigma}^{(e)} + \Hat{\alpha} \textbf{Q} $. In the definition of total stress tensor ($\boldsymbol{\sigma}$), $\mu$ corresponds to the fluid viscosity, $\mathbf{E}$ is the strain-rate tensor ($E_{ij} = \frac{1}{2}(\partial_i v_j + \partial_j v_i)$), $P$ is the pressure and $\boldsymbol{\sigma}^{(e)}$ is the elastic stress tensor. The elastic stress tensor is given as $\sigma_{ij}^{(e)} = -\Hat{\lambda} S H_{ij} + Q_{ik} H_{kj} - H_{ik} Q_{kj}$, where $H_{ij}$ is the molecular tensor and  $\Hat{\lambda}$ is the flow aligning parameter. The molecular tensor is given as $H_{ij} = -\delta F/\delta Q_{ij}$.\\

In a corrugated channel, coherent flow in region I is predominantly unidirectional and thus assuming a translationally invariant flow in the $x$ direction, the flow field is described by the velocity field $v_x = v_x (y) $ and $v_y =0 $. The strain-rate tensor $\mathbf{E}$ has only one no-zero component $E_{xy} = \partial_y v_x$, thus Eq.~\ref{convective} is reduced to the following form:
\begin{equation}
    \rho \partial_t v_x = \partial_y \sigma_{xy}.
\end{equation}
On solving for the total stress tensor $\sigma_{xy}$ and neglecting non-linear terms, and elasticity, the aforementioned equation can be written as:
\begin{equation}
    \rho \partial_t v_x = \partial_y \Bigl( \mu \partial_y v_x + \Hat{\alpha} S \sin 2\theta \Bigr).
    \label{v final}
\end{equation}
The hydrodynamic equation for the nematic tensor order parameter $\mathbf{Q}$ can be written in the form:
\begin{equation}
    [\partial_t + v \cdot \nabla]\mathbf{Q} = \Hat{\lambda} S \mathbf{E} + \mathbf{Q}\cdot\boldsymbol{\Omega} - \boldsymbol{\Omega}\cdot\mathbf{Q} + \gamma^{-1} \mathbf{H}
    \label{Q full equation}
\end{equation}
where, $\boldsymbol{\Omega}$ is vorticity tensor which is given as $\Omega_{ij} =\frac{1}{2}(\partial_i v_j - \partial_j v_i)$. 
Following\cite{Giomi_2012} and decoupling the equation for the  the orientation angle $\theta$ of the nematogens from that of the  the scalar order parameter $S$ we obtain,
\begin{equation}
    \partial_t \theta = \gamma^{-1} K \partial_y^2 \theta - \frac{\partial_y v_x}{4}(2-\Hat{\lambda} \cos 2\theta).
    \label{theta final}
\end{equation}\\
Assuming that the viscous stress dominates at the boundary, and hence using the Newtonian constitutive relation, the boundary conditions can be written as,
\begin{align}
    v_x(y=0~\text{and}~y=W) &= \beta \frac{dv}{dy} \\
    \theta(y=0~\text{and}~y=W) &= \frac{\pi}{2}
\end{align}

For the linear stability analysis, we consider $\varphi(y,t) = \varphi_0 + \epsilon \varphi_1(y,t)$, where, $\varphi = \{\theta, v_x\}$ and $\varphi_0 = \{\pi/2, 0\}$ and $\epsilon \ll 1$. Substituting $\theta$ and $v_x$ in Eq.~\ref{v final} and  Eq.~\ref{theta final} and considering the system at steady state yields the linearised system of equations:
\begin{align}
    \mu \partial_y^2 v_1 - 2\Hat{\alpha} S \partial_y \theta_1 &=0 
    \label{stability 3}\\
    \gamma^{-1} K \partial_y^2 \theta_1 - \frac{\partial_y v_1}{4}(2+ \Hat{\lambda}) &=0
    \label{stability 4}
\end{align}
Solving the coupled, homogeneous differential equations we obtain,
\begin{align}
    \theta_1 &= \frac{C_1}{T} \sin Ty - \frac{C_2}{T} (1-\cos Ty)
    \label{theta sol final}\\
    v_1 &= \frac{4 \gamma^{-1} K}{2+ \Hat{\lambda}} \left(C_1(\cos Ty -1) + C_2 (\sin Ty + \beta T)\right)
    \label{v sol final}
\end{align}
where $C_1$ and $C_2$ are constants to be determined and
\begin{equation}
    T = \sqrt{\frac{\Hat{\alpha}(2+\Hat{\lambda})S}{2\mu \gamma^{-1} K}}.
    \label{root}
\end{equation}

\begin{figure*}[p!t!]
    \includegraphics[width=\textwidth]{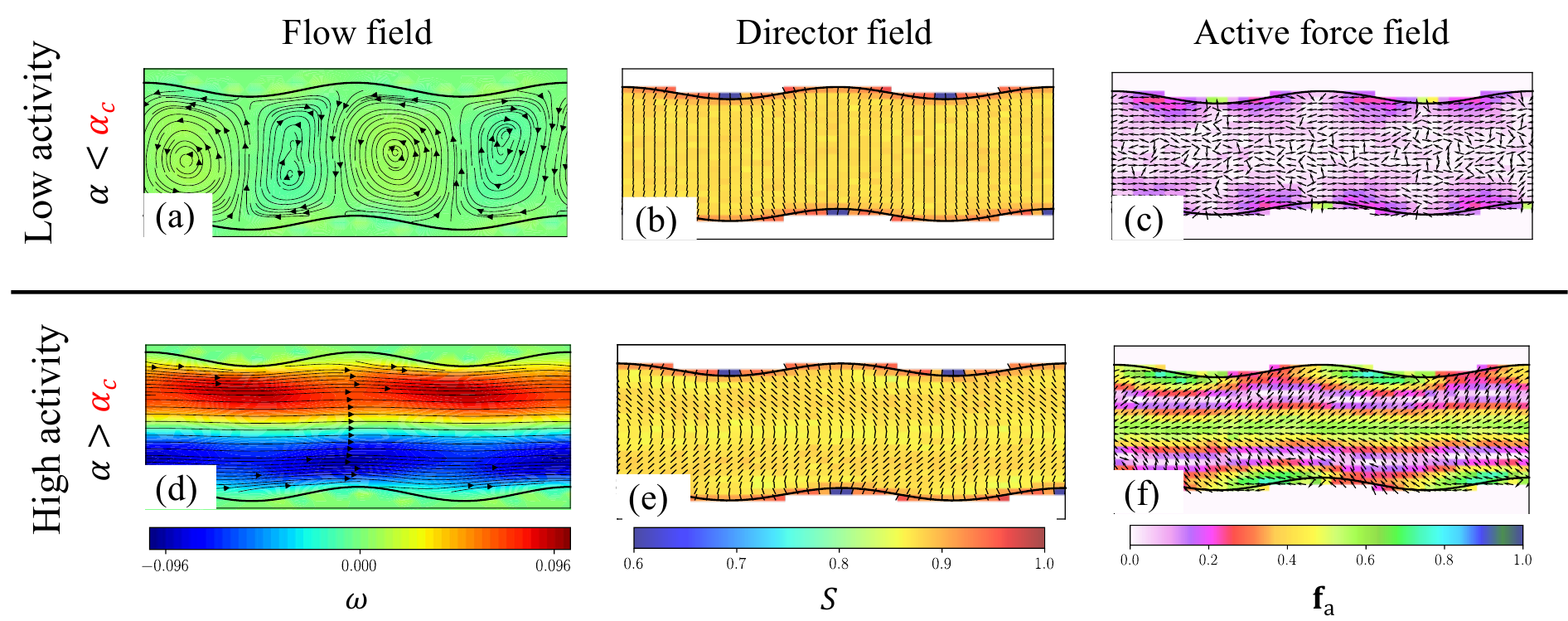}
    \caption{Snapshots of the flow field (left column), director field (middle column) and active forcing (right column) in channels with in-phase corrugated walls. In the left column, the flow field is shown, which is represented by streamlines superimposed on the vorticity of the fluid. The colorbar is shown at the bottom  with blue and red representing clockwise and anticlockwise rotation respectively. In the middle column, the director field is shown by black dashed line color shaded by scalar order parameter. In the right column, the active force is illustrated by black arrows and are color coded by the magnitude of the force. Figures (a)-(c) correspond to  low activity and figures (d)-(f) correspond to higher activity at which coherent flows prevail. All channels have a mean width of $W=20$, amplitude $A=1$ and wavelength $\Lambda=20$.}
    \label{fig:in-phase}
\end{figure*}

Applying the boundary conditions we obtain the condition for the critical activity for spontaneous flow transition:
\begin{equation}
    1 - \cos TW + \frac{\beta}{W} TW \sin TW =0 .
\end{equation}
This equation is further simplified by considering $\varepsilon = \beta/W$ as a small parameter and obtain the critical activity as:
\begin{equation}
    \Hat{\alpha}_{c,s} = \frac{8 \pi^2 \mu \gamma^{-1} K (1-2\varepsilon +4 \varepsilon^2)^2}{W^2 S (2+\Hat{\lambda})}.
    \label{critical activity slip full}
\end{equation}

\section{In-phase corrugated channel}
\label{Appendix D}

Here we present some results of active nematic confined in a corrugated channel, with the top and bottom walls in-phase, unlike the out-of-phase channel walls described in the main manuscript. The boundary conditions are as mentioned in section~\ref{2.2}: homeotropic anchoring for the director field and no-slip for the velocity field on the walls. The channel walls have an amplitude $A=1$, wavelength = $20$ with a mean width of $W=20$.\\
When the activity $\alpha$ is less than critical activity $\alpha_c$, the channel with in-phase walls shows a lattice of swirling flow structures centred on the centreline (Fig.~\ref{fig:in-phase}(a)). Since the curvature of the top and bottom channel walls are same, the anchoring boundary conditions (namely the geometry) will result in variations in the director field primarily along the channel length. The variations are in an anti-symmetric fashion such that the corresponding flow generated will be in opposite directions. Hence, vortices span the width of the channel (Fig.\ref{fig:in-phase}(a) in contrast to Fig~\ref{fgr:wavyvsflat}(d)). It is interesting to note that the structure of the flow is similar to the well-known dancing flow state in confined active nematics, but now with a major difference: here, the lattice structure of the flow vortices arise purely from the geometry without the presence of topological defects\cite{shendruk2017dancing,Santhan2D,Santhan3D,Abhik} (Fig.~\ref{fig:in-phase}(b)).

\begin{figure}
     \centering
     \begin{subfigure}[b]{0.46\textwidth}
         \centering
         \includegraphics[width=\textwidth]{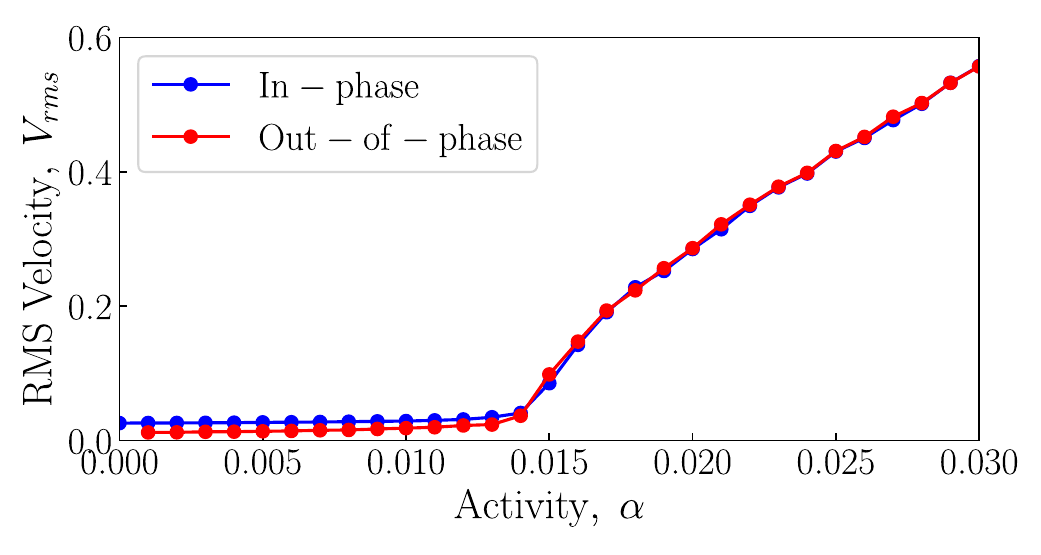}
         \caption{}
     \end{subfigure}
     \hfill
     \begin{subfigure}[b]{0.46\textwidth}
         \centering
         \includegraphics[width=\textwidth]{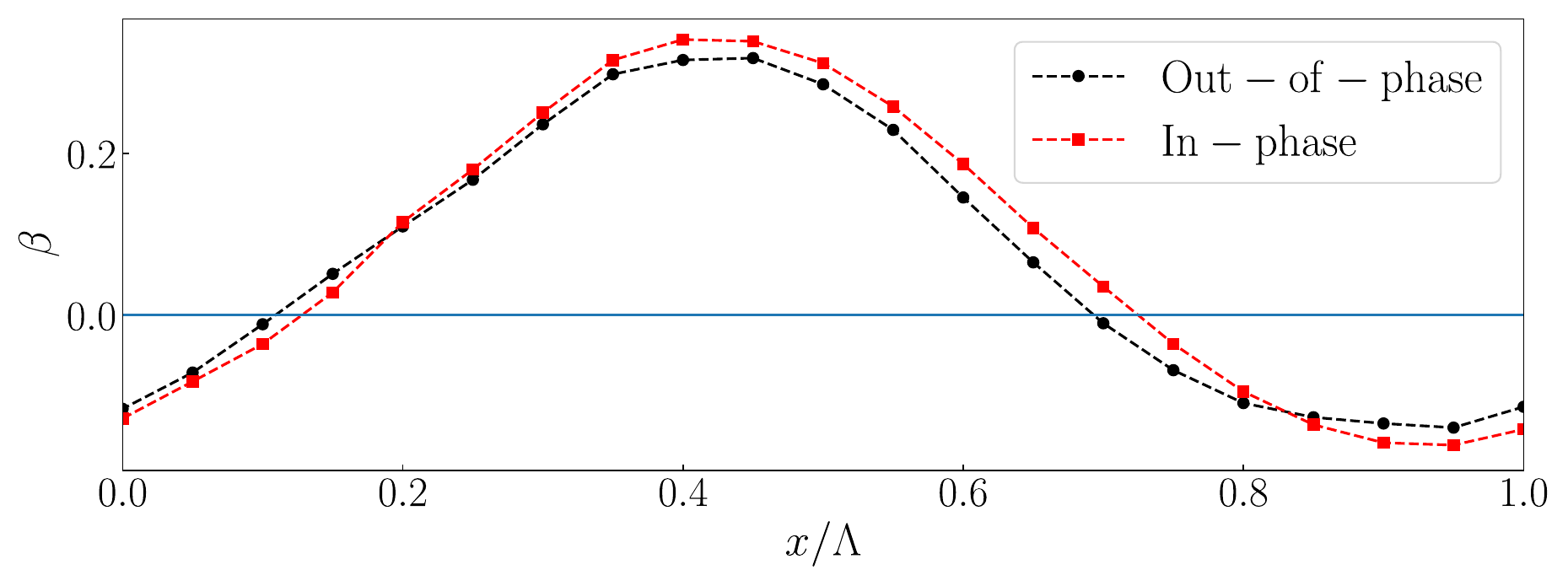}
         \caption{}
     \end{subfigure}
     \caption{(a) Comparison of spontaneous flow transition observed in channels with in-phase and out-of-phase walls by plotting root mean square velocity as a function of activity. (b) Variation in slip length $\beta (x)$ for the in-phase and out-of-phase corrugated channels at $\alpha=0.024$. The parameters used in the simulations are same as those mentioned in the caption of Fig.~\ref{fig:in-phase}.}
     \label{fig:in_vs_out}
\end{figure}

Upon spontaneous flow transition, symmetry along the channel length is broken, and a coherent flow is observed. The sinusoidal variations in the active force field will follow the geometry of the channel with in-phase walls as shown in Fig.~\ref{fig:in-phase}(f), which is unlike the reflection symmetry in the active force field observed in the case of channel with out-of-phase walls (Fig.~\ref{fgr:wavyvsflat}(l)). A comparison of root mean square velocity of the active nematic fluid as a function of activity is shown in Fig.~\ref{fig:in_vs_out}(a) for both in-phase and out-of-phase channels. In accordance with the discussion above, strength of the flow before the flow transition is different in two channels, but the strength of the flow is comparable in the two channels after the flow transition. Fig.~\ref{fig:in_vs_out}(b) shows a comparison of the slip length as a function of channel length in both types of channels. It may be seen that, the magnitude and variation of slip length $\beta$ in both out-of-phase and in-phase channel are similar.
It is clear from the above discussions that the qualitative features and the mechanisms at play in channels with walls of broken symmetry remain same.

\section*{Acknowledgements}
SPT acknowledges the support by Department of Science and Technology, India via the research grant CRG/2023/000169.

\noindent This research has received funding (T.N.S.) from the European Research Council under the European Union’s Horizon 2020 research and innovation programme (Grant Agreement No. 851196). 



\balance


\bibliography{rsc} 
\bibliographystyle{rsc} 

\end{document}